\documentclass[pre,preprint,floatfix]{revtex4}
\usepackage{amsmath,graphicx}
\usepackage{epsfig}
\usepackage{color}
\usepackage{acronym}

\begin{document}

\title{Percolation of particles on recursive lattices:\\
(III) percolation of polydisperse particles in the presence of a polymer matrix}
\author{Andrea Corsi\footnote{Present address: Dipartimento di Fisica, Universit\`a degli Studi di Milano,
Milano, Italy}, P. D. Gujrati\footnote{Electronic address: pdg@physics.uakron.edu}}
\affiliation{The Department of Physics and The Department of Polymer Science, The University of Akron, Akron, Ohio, 44325, U.S.A.}
\date{\today }

\begin{abstract}
We use a recently developed lattice model to study the percolation of particles of different sizes and shapes in the presence of a polymer matrix. The polymer is modeled as an infinitely long chain to simplify the calculation but we make it more realistic by considering it semi-flexible. We study the effects of the stiffness of the polymer, the size disparity of the filler particles, their aspect ratio and the interactions between fillers and polymer on the percolation properties of the system. The lattice model is solved exactly on a recursive square Husimi lattice, which is an approximation for a square lattice. The solution represent an approximate solution for a square lattice. Our results are able to reproduce most of the experimental findings that have been observed in the literature. In particular, we observe how an increase in the size disparity of the filler particles dispersed in the matrix as well as an increase of their aspect ratio decreases the percolation threshold.
\end{abstract}

%\preprint{UATP/06-04}
\maketitle

\section{Introduction}

In the past several years, many studies have been devoted to the analysis of the physical properties of filled polymers. As explained in the previous papers \cite{firstpaper, secondpaper}, polymers are filled because the presence of fillers can improve various physical properties: mechanical and rheological properties can be modified, as well as the electrical resistivity, the absorption of ultraviolet and infrared radiation, the optical clarity, etc. Therefore, the fillers that are introduced in the matrix of a polymer composite serve a useful purpose.

Mechanical properties have been studied more extensively. For example, the addition of different volume fractions of carbon black to a polypropylene (PP) matrix can enhance the Young's modulus of the matrix. The composite that is produced is consequently stiffer than the matrix \cite{Chodak-01}. 
The processing conditions can also affect the properties of a composite. For a fixed fraction of black filler, samples prepared by injection molding have a Young's modulus that is consistently higher than the modulus of samples prepared with compression molding. This effect has been attributed \cite{Chodak-01} to a better distribution of the filler in the PP matrix due to 
additional compounding under shear in the injection-molding equipment. Orientation due to elongational flow during injection molding might also play a role while compression molded samples are supposedly isotropic \cite{Chodak-01}.

The addition of a filler never improves all the properties of a matrix. Usually a decrease in the elongation at break of polymer samples filled with carbon black accompanies an increase in their Young's modulus. The composites that have a high filler content have a very low deformation at break. They break before they are able to achieve a significant degree of orientational reinforcement due to drawing during the tensile tests. The elongation at break of the compression molded samples, as well as their tensile strength \cite{Chodak-01}, is lower than that of the injection molded samples indicating the presence of larger or more frequent defects, like voids or cracks, as a consequence of a less perfect filler dispersion. The addition of a filler to a matrix can improve other properties. If the electrical properties of the same PP composites filled with carbon black are studied one observes a sharp increase in the electrical conductivity and a sharp decrease in the elongation at break, at about the same amount of filler, around 7\% in weight \cite{Chodak-01,Chodak-99}.

The increase of the electrical conductivity and the decrease of the elongation at break are related, and may be associated with the formation of a more or less percolating filler phase in the composite; this filler phase allows for a much higher electrical conductivity on the one hand, but also leads to the formation of more frequent failure sites on the other hand. The value of filler content at which the physical properties of a filled composite change dramatically is considered to be the percolation threshold. In this case, the value of the percolation threshold for the two processes is the same. However, if the changes in two different physical properties are associated with different microscopic mechanisms, the threshold associated with different physical properties should not coincide, as explained in the previous paper \cite{secondpaper}.

The results for the conductivity and the elongation at break of the same composites that have been prepared by injection molding show a much higher percolation threshold, about 16\% in weight \cite{Chodak-01}. This observation is consistent with the idea that because of the different preparation conditions, the carbon black is less perfectly distributed in the PP matrix in the compression molded samples than in the injection molded samples. This allows for the formation of conducting channels formed by carbon black particles at much lower concentrations. As percolation is expected to be affected by size variations of filler particle, it is important to quantitatively assess their effects on percolation. The particles' size, shape and polydispersity after the composite has been processed matters most because these properties can be greatly affected by the processing itself \cite{Gokturk-93}. Our results here will show that an increased polydispersity of the filler particles may be responsible for the reduction of the percolation threshold, and has been observed also in other works \cite{Gokturk-93}.

Although polymers are good electric insulators, they are usually poor thermal conductors and their dielectric properties are strongly affected by the presence of moisture. One solution to this problem is to make polymer composites with increased thermal conductivity. This has been achieved by adding proper fillers, characterized by a high thermal conductivity, to the polymer matrix. The thermal conductivity increases as the volume fraction of the filler increases in all cases, as predicted by many theories and observed in experimental studies \cite{Lu-97,Hatta-92,Bigg-95,Hill-02,Danes-03}. Carbon fibers increase the  thermal conductivity of the polyurethane matrix more than the alumina particles. This is certainly related to the intrinsic properties of the reinforcing particles, the thermal conductivity of carbon black being higher than that of alumina. Also, both an increase in the size of the alumina particles and a decrease in the aspect ratio of the carbon fibers decreases the thermal conductivity. From the point of view of percolation, the decreased thermal conductivity should correspond to an increase in the value of the volume fraction occupied by the filler at the percolation threshold. We have here another experimental evidence of how the size of the particles can affect the percolation threshold of a composite system. In general, an increase of the size of the particles increases the volume fraction of the filler at the percolation threshold. A decrease of the aspect ratio of the filler particles has a similar effect.

The shape of the particles has also a critical effect on the electrical properties of the composite. Higher aspect ratios facilitate the formation of clusters and therefore lower the threshold concentration \cite{Bigg-84}. The dielectric constant of composites samples containing asymmetric fillers usually exhibit a strong dependence on the frequency, increasing significantly as the frequency decreases. This significant enhancement of the dielectric constant values at low frequency can also be explained by the formation of filler clusters in the composites \cite{Gokturk-93}. At low frequencies, the surface charges that contribute to the electrical polarization are given more time to rearrange themselves on the clusters that have larger dimensions, generating greater polarizations. It is important to observe that in many experimental works, a change in particle shape can be associated with an increase in the polydispersity of the particles so that the observed increase in the dielectric constant, and consequently the decrease in the percolation threshold, could be ascribed to an increase in the polydispersity of the filler particles, as well as to a combination of the two effects. The addition of 20\% volume of nickel to a polyethylene (PE) matrix affects the dielectric properties of the polymer matrix \cite{Gokturk-93}. As expected, the dielectric constant values increase as the filler is added to the polymer matrix. The increase in the  dielectric constant in the presence of metallic fillers has been explained as due to electrical polarization of the charges induced on the surfaces of  the metallic filler particles \cite{Gokturk-93}. The dielectric constant of composites with filaments and flake-like fillers is higher than the one of the composites filled with nickel powder and increases with increasing aspect ratio. The enhancement of the dielectric constant at higher aspect ratios has been explained as due to an increased easiness of formation of clusters of filler particles \cite{Gokturk-93}. 

The presence of filler particles, and their nature, can also affect the  rheological properties of polymers. In general, the addition of a filler increases the viscosity of the polymer matrix \cite{Adams-93,Adams-Clay,Ghosh-00}. Such increase is usually proportional to the fraction of filler present in the system until a critical value is reached where a sharp jump corresponding to the percolation threshold occurs. The limiting shear viscosity, as well as the yield stress, of carbon black filled ethylene propylene diene monomer (EPDM) system composites depends on the fraction of black. The sudden increase in the shear viscosity around a value of 15phr (parts per hundred rubber) of carbon black, is explained in terms of the development of a trend in filler particle networking that becomes very prominent at what is thought to be the percolation threshold  \cite{Wolff-92,Wang-Wolff-92,Medalia-78}. The incorporation of fillers also enhances the low shear rate melt viscosity of EPDM. The narrowing of the difference in the shear viscosity values between compounds of different filler loadings at high shear rates occurs as a result of a progressive breakdown of the filler network structure \cite{White-74}. As mentioned above, it is very complicated to measure and model the viscoelastic properties of composite systems since these properties depend on the dynamics of the system. The shearing rate at which the viscoelastic properties are measured, for example, is extremely important since high shear rate tend to largely nullify the effects of filler networking and the effect of filler-polymer physical interactions \cite{Ghosh-00}. In general, decreasing the size and the aspect ratio of fillers increases the viscosity  \cite{Adams-Clay}.

There is enough experimental evidence to conclude that the properties of the composite are affected by a large number of parameters: the size, shape and distribution of the reinforcing particles as well as the interactions between the particles and the polymeric matrix. It is often difficult, if not impossible, to separate the effects due to the different parameters. To summarize the experimental results, we can say that the following general trends have been observed:

\begin{itemize}
\item the percolation threshold appears to decrease as the size disparity between different particles present in the system increases,  

\item the percolation threshold appears to decrease with the aspect ratio of  the particles present in the system,

\item the nature of the interactions between fillers and matrix affects the percolation properties of the system.
\end{itemize}

\section{Percolation of single-site filler particles in the presence of a polymer matrix}

\subsection{Model}

In order to describe the percolation of particles of different sizes in the presence of a polymer matrix, we have used a model for a \emph{semiflexible} polymer due to Flory \cite{Flory-42,Flory-56}, which has been extended and used previously in our research group \cite{PDG-Corsi-PRL-01,PDG-Corsi-PRE-03,PDG-Rane-Corsi-PRE-03,Rane-PDG-Macro-05}.
This polymer model, originally used to describe an infinite molecular weight incompressible polymer, has been recently implemented to take into account the finite length of many polymers and the presence of free volume \cite{PDG-Rane-Corsi-PRE-03,Rane-PDG-Macro-05}. The model developed in \cite{PDG-Corsi-PRE-03} is defined on a square lattice and solved \emph{exactly}
on a square Husimi lattice, part of which is shown in Figure \ref{Husimi}. The Husimi lattice is built in a recursive fashion starting from a central plaquette. The different size of the plaquettes that belong to different generations is just a matter of convention: one might consider all plaquettes to have the same size but then a lattice would become impossible to draw because of the crowding that one has as one moves away from the origin of the lattice. 

In the present research, we extend this model.  We consider a system made of a semiflexible polymer matrix, in the limit of infinitely long polymer chains, and filler particles of different sizes embedded in such a polymer matrix. Each monomer belonging to the polymers occupies a lattice site, and the lattice bond connecting two occupied nearest-neighbor sites represents a polymer bond in the polymer. The number density $\phi _{\text{P}}$ of infinitely long polymer chains is kept equal to zero. In the following, we will often loosely speak of a single polymer, which also corresponds to $\phi _{\text{P}}=0,$ to describe this limit, even though the limit only ensures that the number of polymers does not grow as rapidly as the system size $N$, where $N$ denotes the number of lattice sites. In this limit, either there exist no polymers if their monomer density is zero or they exist if their monomer density is non-zero. It is possible in the latter case to have polymers cover a finite non-zero fraction of the entire lattice. This is the reason why we can have configurations in which two polymer bonds are parallel to each other within a plaquette without being connected to each other, see below. As explained elsewhere \cite{PDG-Corsi-PRE-03}, the limit of infinitely long polymers ($\phi _{\text{P}}=0$) corresponds to a situation in which the end-points, whose density ($=2\phi_{\text{P}}$) is also zero, do not have any effect on the thermodynamics of the system.
As a consequence one might think of this limit as useful for chains that are long enough so that the end points represent a negligible fraction of the polymeric monomers present in the system. 

Keeping $\phi _{\text{P}}=0$ fixed throughout the study (see below to see how we ensure $\phi _{\text{P}}=0)$, we have studied the percolation properties of systems containing up to two different kinds of filler particles. First, we analyzed the percolation of single-site particles which occupy one single site each on the lattice, just as a polymeric monomer. In the following, the single-site filler particles are going to be called solvent. To simplify the lattice model, we take the solvent and the polymeric monomers to have the same size. This restriction can be easily removed at the expense of making the recursive calculation more cumbersome. This will be done elsewhere. Then, in the following sections, we study the percolation of polydisperse fillers made of single-site solvent particles and square-shaped particles in one case and single-site solvent particles and star-shaped particles in the other. 

The original model used in the literature to describe the thermodynamics of semiflexible polymer chains was developed originally independently by Flory \cite{Flory-42,Flory-56} and Huggins \cite{Huggins-42}. This model was the basis for the Gibbs-DiMarzio theory of the glass transition that has represented for many years one of the most successful theories used to describe the glass transition in polymers \cite{Gibbs-58-I,Gibbs-58-II,Gibbs-63}. Unfortunately the approximations made in this theory have been proved to be qualitatively wrong \cite{PDG-80,PDG-Goldstein-81,PDG-82}, even though the resulting physics behind the entropy crisis seems to be valid not only for long polymers but also for smaller molecules, as demonstrated recently by our group \cite{PDG-Corsi-PRL-01, PDG-Corsi-PRE-03, Semerianov-03}. This has been one of the driving forces towards developing a new theory. Our current model is an extension of Flory's model and his model is recovered in a particular limit  \cite{PDG-Corsi-PRE-03}.

In our model, a polymer chain is assumed to consist of $n$ equal monomers, each of the same size as the solvent molecule, which we always designate as the smallest among all the filler particles which are present in the system. For simplicity (see above), we consider the solvent to occupy a single lattice site. This condition can be easily relaxed, but we will not do that here. Any larger filler particle is assumed to contain $n^{\prime }>1$ contiguous monomers so that it covers $n^{\prime }$ contiguous sites on the lattice. Each site of the lattice is occupied by either a polymer chain or a filler particle, and the excluded volume effects are taken into account by requiring a site to be occupied only once. The polymer chain occupies a contiguous sequence of all the lattice sites connected by polymer bonds, just as is the case of a filler particle with $n^{\prime }>1,$ 
 without forming any branches or loops. 

For concreteness and ease of discussion, one can think of the system as if it were defined on a square lattice. At every site which is occupied by a monomer, the polymer chain can assume either a trans conformation (the conformation is related to the state of two consecutive bonds, and, consequently, of three consecutive monomers, when the consecutive bonds are collinear, or one of the two possible gauche conformations, when the polymer chain bends. For a semiflexible polymer chain, every gauche conformation has an energy penalty $\varepsilon $\thinspace compared to a trans conformation. This is the only energy present in the original model developed by Flory. We set the energy for a trans conformation to be zero. The inclusion of this single energy for a gauche bond is not satisfactory since it leads to a continuous melting transition that does not occur in nature \cite{PDG-Corsi-PRE-03}. In order to obtain a meaningful description of the physics of semiflexible polymer chains, it is necessary to include other conformational energies that were not included in Flory's model. We introduce two such energies: an interaction energy $\varepsilon _{\mathrm{p}}$ associated with each parallel pair of neighboring bonds, and an energy $\varepsilon _{\mathrm{h}}$ for each hairpin turn within each square of the Husimi lattice \cite{PDG-Corsi-PRE-03}. Corresponding to each quantity $N_{\mathrm{g}}$, $N_{\mathrm{p}}$ and $N_{\mathrm{h}}$, the number of gauche conformations, parallel bond pairs and hairpin turns, respectively, there is an independent activity $w$, $w_{\mathrm{p}}$ and $w_{\mathrm{h}}$, respectively. The activities $w$, $w_{\mathrm{p}}$ and $w_{\mathrm{h}}$ are defined as follows: 
\begin{equation}
w\equiv \exp \left( -\beta \varepsilon \right) ,\ w_{\mathrm{p}}\equiv \exp
\left( -\beta \varepsilon _{\mathrm{p}}\right) \equiv w^{a},\ w_{\mathrm{h}
}\equiv \exp (-\beta \varepsilon _{\mathrm{h}})=w^{b}.
\end{equation}
Here, $\beta $ is the inverse temperature \textit{T} in the units of the Boltzmann constant, and $a=\varepsilon _{\mathrm{p}}/\varepsilon $, $b=\varepsilon _{\text{h}}/\varepsilon $. The original model developed by Flory is obtained when the last two interactions are absent, so that $w_{\mathrm{p}}=w_{\mathrm{h}}=1$. It should be stressed that $\varepsilon _{\mathrm{h}}$ is the additional energy associated with the conformation, once the energy of the two bends and a pair of parallel bonds have been subtracted out. Both $\varepsilon _{\mathrm{p}}$ and $\varepsilon _{\mathrm{h}}$\ are  associated with four-site interactions, since it is necessary to determine the state of four adjacent sites to determine if a pair of parallel bonds or a hairpin turn is present. It has been seen that $a$ and $b$ represent possible measures of the stiffness of the polymer chain \cite{Rane-PDG-Macro-05}.

In the presence of single-site solvent particles, the total partition function of the system can be written as 
\begin{equation}
Z_{N}\equiv \sum \eta ^{N_{\mathrm{S}}}H^{2P}w_{\text{c}}^{N_{\text{c}}}w^{N_{\mathrm{%
g}}}w_{\mathrm{p}}^{N_{\mathrm{p}}}w_{\mathrm{h}}^{N_{\mathrm{h}}},
\label{PF1}
\end{equation}%
where the number of solvent molecules $N_{S}$ is controlled by the solvent activity $\eta =w^{-\mu }$, where $\mu $ is the chemical potential of the filler particle. The number $P$ of polymers is controlled by another  activity given by $H^{2}.$ The interaction between nearest neighbor pairs $N_{\text{c}}$ of filler particles and monomers of the polymer is given by the Boltzmann weight $w_{\text{c}}=\exp \left( -\beta \varepsilon _{\mathrm{c}}\right) =w^{c}$. In all the cases studied during this research, the polymers have been considered of infinite length, so that $H=0$.

\subsection{States of a site}

As the model cannot be solved exactly on any regular lattice, we replace the original square lattice with a Husimi lattice \cite{firstpaper}. We consider a linear polymer that covers all the sites of the Husimi lattice along with solvent molecules which occupy one site each. In order to be able to describe all the phases present in the phase diagram of the system, it is necessary to use a labeling scheme in which, if the base site of the plaquette has the label $m$, the two middle sites have the label $(m+1)$ and the peak site has the label $(m+2)$. This is a slightly different labeling than the original one used in the previous paper, where all the sites, other than the base one, were labeled as $(m+1)$ sites. As explained in \cite{secondpaper}, the difference between the two labeling schemes is related to the different scheme adopted for the solution of the equations that we need to solve in order to obtain the thermodynamics of the system. In this case we are going to use a two-cycle recursion scheme so that the current labeling becomes fundamental to properly describe the solutions.

Consider an $m$th level square. The base site can assume five different possible states depending on the state of its lower bonds connected to the base site, as shown in Figure \ref{Figure: Polymer+solvent states}:

\begin{enumerate}
\item In the $\text{I}$ state, both lower bonds are occupied by the polymer chain. Since the polymer is linear, the two upper bonds in the lower $(m-1)$th level square must be unoccupied by the polymer.

\item In the $\text{O}$ state, both lower bonds are unoccupied by the polymer chain but the upper bonds in the $(m-1)$th level square are occupied so that a monomer belonging to a polymer chain occupies the base site of the plaquette under investigation.

\item In the $\text{L}$ state, only one of the lower bonds is occupied and the polymer occupies the left upper bond of the lower $(m-1)$th level square (we always think about left and right moving down the lattice towards its center).

\item In the $\text{R}$ state, only one of the lower bonds is occupied and the polymer occupies the right upper bond of the lower $(m-1)$th level square. 

\item In the $\text{S}$ state, the base site of the plaquette under investigation is occupied by a solvent particle.
\end{enumerate}

For $m=0$, the lower square in the above classification is the square on the other side of the origin.

We are interested \cite{firstpaper} in the contribution of the portion of the $m$th branch $\mathcal{C}_{m}$ of the lattice to the total partition function of the system. In order to carefully account for statistical
weights, a Boltzmann weight equal to $w$ is considered only if the bend
happens:

\begin{enumerate}
\item At the $m$th level and at least one polymer bond at the level is inside the square.

\item At the $(m+1)$th or $(m+2)$th level, and both polymer bonds at the level are inside the polymer.
\end{enumerate}

A weight $w_{\mathrm{p}}=w^{a}$ is considered for any configuration in which two bonds are parallel to each other within the same square. We can, furthermore, distinguish configurations in which two bonds are parallel to each other from configurations where three consecutive bonds form a hairpin configuration. As explained above, whenever this configuration is present, an additional weight $w_{\mathrm{h}}=w^{b}$ is introduced.

It is not important to know along which of the two lower bonds in the $(m+1)$th or $(m+2)$th square does the polymer chain enter into the $m$th square. In fact, even if the polymer undergoes a bend while moving from the higher level square to the $m$th level square the corresponding weight is already taken into account into the partial partition function of the higher level site.

A weight $w_{\mathrm{c}}=w^{c}$ is considered for any configuration in which a filler particle and a monomer belonging to a polymer chain are nearest neighbors. In order to consider two particles in contact we need them to be nearest neighbors on the lattice. In other words, it must be possible to reach one particle from the other by moving one single step on the lattice. For example, the base and the peak sites inside one plaquette are not nearest neighbors because there is no single lattice bond connecting them. The same is true for the two middle sites of the square. The base site or the peak site is a nearest neighbor of the two middle sites.

\subsection{Recursion Relations}

Following the usual procedure explained in Ref. \cite{firstpaper}, the next step consists in constructing the recursion relations (RRs) for the partial partition functions. Contrary to Refs. \cite{firstpaper} and \cite{secondpaper}, we start right away with constructing the RRs for a particular 2-cycle solution, suitable for describing the ground state (at absolute zero), in which the base site of the plaquette under investigation has label $m$, but we make a distinction between the middle sites and the peak site so that the middle sites have label $m+1$ and the top site has the label $m+2$. There is no harm in doing this as the 1-cycle solution is a special case of this scheme as noted in the previous paper of the series, and as we will further clarify below. In other words, the scheme will produce a 1-cycle solution that will describe the disordered phase at high temperatures. Therefore, it is a more general solution that a 1-cycle solution.

The recursion relation for $Z_{m}(\text{I})$, the partial partition function of the $m$th branch of the Husimi lattice given that the $m$th level site is in the $\text{I}$ state, is written as a function of the $Z_{m+1}(\alpha )$ and $Z_{m+2}(\beta )$, the partial partition functions describing the contributions from the upper branches and summing up all the contributions coming from all the possible configurations that are compatible with the state at the $m$th level site. 
\begin{align}
Z_{m}(\text{I})& =w^{2}w_{\mathrm{h}}w_{\mathrm{p}}Z_{m+1}(\text{O})[Z_{m+1}(%
\text{L})Z_{m+2}(\text{R})+Z_{m+1}(\text{R})Z_{m+2}(\text{L})]  \notag \\
& +wZ_{m+1}(\text{R})Z_{m+1}(\text{L})(Z_{m+2}(\text{I})+w_{\mathrm{c}%
}^{2}Z_{m+2}(\text{S})).
\end{align}%
Considering the case in which the $m$th level site is in the $\text{O}$ state, the partial partition function $Z_{m}(\text{O})$ for the $\text{O}$ state can be written as: 
\begin{align}
Z_{m}(\text{O})& =Z_{m+1}^{2}(\text{I})Z_{m+2}(\text{I})+2w_{\mathrm{c}%
}^{2}Z_{m+1}(\text{I})Z_{m+2}(\text{I})Z_{m+1}(\text{S})+w_{\mathrm{c}%
}^{2}Z_{m+1}^{2}(\text{I})Z_{m+2}(\text{S})  \notag \\
& +2w_{\mathrm{c}}^{2}Z_{m+1}(\text{I})Z_{m+2}(\text{S})Z_{m+1}(\text{S})+w_{%
\mathrm{c}}^{4}Z_{m+1}^{2}(\text{S})Z_{m+2}(\text{I})+w_{\mathrm{c}%
}^{2}Z_{m+1}^{2}(\text{S})Z_{m+2}(\text{S})  \notag \\
& +[Z_{m+1}(\text{I})+w_{\mathrm{c}}^{2}Z_{m+1}(\text{S})]\left[ Z_{m+1}(%
\text{L})Z_{m+2}(\text{R})+Z_{m+1}(\text{R})Z_{m+2}(\text{L})\right]  \notag
\\
& +wZ_{m+1}(\text{R})Z_{m+1}(\text{L})Z_{m+2}(\text{O}).
\end{align}%
When the $m$th level site is in the $\text{L}$ state the partial partition function can be written as: 
\begin{align}
Z_{m}(\text{L})& =[Z_{m+1}(\text{R})+wZ_{m+1}(\text{L})]\{Z_{m+1}(\text{I}%
)Z_{m+2}(\text{I})+w_{\mathrm{c}}^{2}Z_{m+1}(\text{S})Z_{m+2}(\text{I}) 
\notag \\
& +w_{\mathrm{c}}^{2}Z_{m+1}(\text{I})Z_{m+2}(\text{S})+w_{\mathrm{c}%
}^{2}Z_{m+1}(\text{S})Z_{m+2}(\text{S})+w^{2}w_{\mathrm{p}}w_{\mathrm{h}%
}Z_{m+1}(\text{O})Z_{m+2}(\text{O})\}  \notag \\
& +w_{\mathrm{p}}\left[ wZ_{m+1}^{2}(\text{L})Z_{m+2}(\text{R})+Z_{m+1}^{2}(%
\text{R})Z_{m+2}(\text{L})\right]  \notag \\
& +\left[ Z_{m+2}(\text{R})+wZ_{m+2}(\text{L})\right] wZ_{m+1}(\text{I}%
)Z_{m+1}(\text{O}).
\end{align}%
The relation for the $\text{R}$ state is easily obtained from $Z_{m}(\text{L})$ by the interchange $L\longleftrightarrow R$: 
\begin{align}
Z_{m}(\text{R})& =[Z_{m+1}(\text{L})+wZ_{m+1}(\text{R})]\{Z_{m+1}(\text{I}%
)Z_{m+2}(\text{I})+w_{\mathrm{c}}^{2}Z_{m+1}(\text{S})Z_{m+2}(\text{I}) 
\notag \\
& +w_{\mathrm{c}}^{2}Z_{m+1}(\text{I})Z_{m+2}(\text{S})+w_{\mathrm{c}%
}^{2}Z_{m+1}(\text{S})Z_{m+2}(\text{S})+w^{2}w_{\mathrm{p}}w_{\mathrm{h}%
}Z_{m+1}(\text{O})Z_{m+2}(\text{O})\}  \notag \\
& +w_{\mathrm{p}}\left[ wZ_{m+1}^{2}(\text{R})Z_{m+2}(\text{L})+Z_{m+1}^{2}(%
\text{L})Z_{m+2}(\text{R})\right]  \notag \\
& +\left[ Z_{m+2}(\text{L})+wZ_{m+2}(\text{R})\right] wZ_{m+1}(\text{I}%
)Z_{m+1}(\text{O}).
\end{align}

Finally, the recursion relation for $Z_{m}(\text{S})$, the partition function of the $m$th branch of the Husimi lattice given that the $m$th level site is in the S state, is given by: 
\begin{align}
Z_{m}(\text{S})& =\eta \lbrack w_{\mathrm{c}}^{2}Z_{m+1}^{2}(\text{I}%
)Z_{m+2}(\text{I})+2w_{\mathrm{c}}^{2}Z_{m+1}(\text{I})Z_{m+2}(\text{I}%
)Z_{m+1}(\text{S})+w_{\mathrm{c}}^{4}Z_{m+1}^{2}(\text{I})Z_{m+2}(\text{S}) 
\notag \\
& +2w_{\mathrm{c}}^{2}Z_{m+1}(\text{I})Z_{m+2}(\text{S})Z_{m+1}(\text{S})+w_{%
\mathrm{c}}^{2}Z_{m+1}^{2}(\text{S})Z_{m+2}(\text{I})+Z_{m+1}^{2}(\text{S}%
)Z_{m+2}(\text{S})  \notag \\
& +[w_{\mathrm{c}}^{2}Z_{m+1}(\text{I})+w_{\mathrm{c}}^{2}Z_{m+1}(\text{S})]%
\left[ Z_{m+1}(\text{L})Z_{m+2}(\text{R})+Z_{m+1}(\text{R})Z_{m+2}(\text{L})%
\right]  \notag \\
& +ww_{\mathrm{c}}^{2}Z_{m+1}(\text{R})Z_{m+1}(\text{L})Z_{m+2}(\text{O})].
\end{align}%
It is possible to write analogous relations for $Z_{m+1}(\mathrm{\alpha })$ by properly substituting $m\rightarrow m+1$, $m+1\rightarrow m+2$ and $m+2\rightarrow m+3.$

We introduce $B_{m}=\left( Z_{m}(\text{L})+Z_{m}(\text{R})\right) $ and the following ratios between partial partition functions at the mth level site of the lattice: 
\begin{align}
x_{m}(\alpha )& \equiv Z_{m}(\alpha )/B_{m},\ \ {\alpha =}\text{ I,O,L, and S%
}  \notag \\
x_{m}(\text{R})& \equiv 1-x_{m}(\text{L}).
\end{align}%
We divide the partial partition functions by $B_{m}$ since we are interested in the limit in which at low temperature the entire lattice is covered by the polymer. So, $B_{m}$ is never zero, and the ratios remain finite at every temperature.

As we climb down any of the branches of the Husimi lattice, a site can be classified as a simultaneous base site (of the upper square) and peak site (of the lower square). These sites differ in their levels (that is, the lattice generation they belong to) by two. Thus, it should come as no surprise that as one moves from a level that is infinitely far away from the origin towards the origin itself, the recursion relations can approach 2-cycle fix-point (FP) solutions, $x_{2m}(\alpha )\rightarrow \alpha _{0}$, $x_{2m+1}(\alpha )\rightarrow \alpha _{1}$, etc., where $\alpha =$ $\text{I}$, $\text{O}$, $\text{L}$,$\text{R}$ or $\text{S}$, regardless of the
initial values of the ratios at infinity. The indexes $0$ and $1$ refer to even and odd levels, respectively. At the fixed point, the values of $\alpha _{0}$ and $\alpha _{1}$ become independent of $m$. Thus, these fix-point solutions of the recursion relations describe the behavior in the interior of the Husimi tree. We express the fix point in this case as follows, where $m$ is taken to be an even integer including $0$: 
\begin{align}
x_{m}(\mathrm{I})& =x_{m+2}(\mathrm{I})=i_{\mathrm{0}},\;x_{m}(\mathrm{O}%
)=x_{m+2}(\mathrm{O})=o_{\mathrm{0}},\;x_{m}(\mathrm{L})=x_{2m+2}(\mathrm{L}%
)=l_{\mathrm{0}},  \notag \\
x_{m}(\mathrm{R})& =x_{m+2}(\mathrm{R})=1-x_{m}(\mathrm{L})=1-l_{\mathrm{0}%
},\;x_{m}(\mathrm{S})=x_{m+2}(\mathrm{S})=s_{\mathrm{0}}  \notag \\
x_{m+1}(\mathrm{I})& =x_{m+3}(\mathrm{I})=i_{\mathrm{1}},\;x_{m+1}(\mathrm{O}%
)=x_{m+3}(\mathrm{O})=o_{\mathrm{1}},\;x_{m+1}(\mathrm{L})=x_{m+3}(\mathrm{L}%
)=l_{\mathrm{1}},  \notag \\
x_{m+1}(\mathrm{R})& =x_{m+3}(\mathrm{R})=1-x_{m+1}(\mathrm{L})=1-l_{\mathrm{%
1}},\;x_{m+1}(\mathrm{S})=x_{m+3}(\mathrm{S})=s_{\mathrm{1}}.
\end{align}

The system of equations for the ratios can be obtained from the system of equations for the partial partition functions and it is easy to show that they can be written in the following form: 
\begin{align}
i_{\mathrm{0}}Q& =w\left( 1-l_{\mathrm{1}}\right) l_{\mathrm{1}}i_{\mathrm{0}%
}+w^{2}w_{\mathrm{p}}w_{\mathrm{h}}o_{\mathrm{1}}\left[ l_{\mathrm{1}}(1-l_{%
\mathrm{0}})+l_{\mathrm{0}}(1-l_{\mathrm{1}})\right] , \\
o_{\mathrm{0}}Q& =i_{\mathrm{1}}^{2}i_{\mathrm{0}}+2w_{\mathrm{c}}^{2}s_{%
\mathrm{1}}i_{\mathrm{1}}i_{\mathrm{0}}+w_{\mathrm{c}}^{2}s_{\mathrm{0}}i_{%
\mathrm{1}}^{2}+2w_{\mathrm{c}}^{2}s_{\mathrm{1}}i_{\mathrm{1}}s_{\mathrm{0}%
}+w_{\mathrm{c}}^{4}s_{\mathrm{1}}^{2}i_{\mathrm{0}}+w_{\mathrm{c}}^{2}s_{%
\mathrm{1}}^{2}s_{\mathrm{0}}  \notag \\
& +i_{\mathrm{1}}\left[ l_{\mathrm{1}}(1-l_{\mathrm{0}})+l_{\mathrm{0}}(1-l_{%
\mathrm{1}})\right] +w(1-l_{\mathrm{1}})l_{\mathrm{1}}o_{\mathrm{0}}, \\
l_{\mathrm{0}}Q& =(1-l_{\mathrm{1}}+wl_{\mathrm{1}})[i_{\mathrm{1}}i_{%
\mathrm{0}}+w_{\mathrm{c}}^{2}s_{\mathrm{1}}i_{\mathrm{0}}+w_{\mathrm{c}%
}^{2}i_{\mathrm{1}}s_{\mathrm{0}}+w_{\mathrm{c}}^{2}s_{\mathrm{1}}s_{\mathrm{%
a}}+w^{2}w_{\mathrm{p}}w_{\mathrm{h}}o_{\mathrm{1}}o_{\mathrm{0}}]  \notag \\
& +wo_{\mathrm{1}}i_{\mathrm{1}}(1-l_{\mathrm{0}}+wl_{\mathrm{0}})+w_{%
\mathrm{p}}(1-l_{\mathrm{1}})^{2}l_{\mathrm{0}}+ww_{\mathrm{p}}l_{\mathrm{1}%
}^{2}(1-l_{\mathrm{0}}),  \notag \\
s_{\mathrm{0}}Q& =\eta \lbrack w_{\mathrm{c}}^{2}i_{\mathrm{1}}^{2}i_{%
\mathrm{0}}+2w_{\mathrm{c}}^{2}s_{\mathrm{1}}i_{\mathrm{1}}i_{\mathrm{0}}+w_{%
\mathrm{c}}^{4}s_{\mathrm{0}}i_{\mathrm{1}}^{2}+2w_{\mathrm{c}}^{2}s_{%
\mathrm{1}}i_{\mathrm{1}}s_{\mathrm{0}}+w_{\mathrm{c}}^{2}s_{\mathrm{1}%
}^{2}i_{\mathrm{0}}+s_{\mathrm{1}}^{2}s_{\mathrm{0}}  \notag \\
& +[w_{\mathrm{c}}^{2}i_{\mathrm{1}}+w_{\mathrm{c}}^{2}s_{\mathrm{1}}]\left[
l_{\mathrm{1}}(1-l_{\mathrm{0}})+l_{\mathrm{0}}(1-l_{\mathrm{1}})\right]
+ww_{\mathrm{c}}^{2}(1-l_{\mathrm{1}})l_{\mathrm{1}}o_{\mathrm{0}}, \\
i_{\mathrm{1}}Q^{\prime }& =w\left( 1-l_{\mathrm{0}}\right) l_{\mathrm{0}}i_{%
\mathrm{1}}+w^{2}w_{\mathrm{p}}w_{\mathrm{h}}o_{\mathrm{0}}\left[ l_{\mathrm{%
a}}(1-l_{\mathrm{1}})+l_{\mathrm{1}}(1-l_{\mathrm{0}})\right] , \\
o_{\mathrm{1}}Q^{\prime }& =i_{\mathrm{0}}^{2}i_{\mathrm{1}}+2w_{\mathrm{c}%
}^{2}s_{\mathrm{0}}i_{\mathrm{0}}i_{\mathrm{1}}+w_{\mathrm{c}}^{2}s_{\mathrm{%
b}}i_{\mathrm{0}}^{2}+2w_{\mathrm{c}}^{2}s_{\mathrm{0}}i_{\mathrm{0}}s_{%
\mathrm{1}}+w_{\mathrm{c}}^{4}s_{\mathrm{0}}^{2}i_{\mathrm{1}}+w_{\mathrm{c}%
}^{2}s_{\mathrm{0}}^{2}s_{\mathrm{1}}  \notag \\
& +i_{\mathrm{0}}\left[ l_{\mathrm{0}}(1-l_{\mathrm{1}})+l_{\mathrm{1}}(1-l_{%
\mathrm{0}})\right] +w(1-l_{\mathrm{0}})l_{\mathrm{0}}o_{\mathrm{1}}, \\
l_{\mathrm{1}}Q^{\prime }& =(1-l_{\mathrm{0}}+wl_{\mathrm{0}})[i_{\mathrm{0}%
}i_{\mathrm{1}}+w_{\mathrm{c}}^{2}s_{\mathrm{0}}i_{\mathrm{1}}+w_{\mathrm{c}%
}^{2}i_{\mathrm{0}}s_{\mathrm{1}}+w_{\mathrm{c}}^{2}s_{\mathrm{0}}s_{\mathrm{%
b}}+w^{2}w_{\mathrm{p}}w_{\mathrm{h}}o_{\mathrm{0}}o_{\mathrm{1}}]  \notag \\
& +wo_{\mathrm{0}}i_{\mathrm{0}}(1-l_{\mathrm{1}}+wl_{\mathrm{1}})+w_{%
\mathrm{p}}(1-l_{\mathrm{0}})^{2}l_{\mathrm{1}}+ww_{\mathrm{p}}l_{\mathrm{0}%
}^{2}(1-l_{\mathrm{1}}), \\
s_{\mathrm{1}}Q^{\prime }& =\eta \lbrack w_{\mathrm{c}}^{2}i_{\mathrm{0}%
}^{2}i_{\mathrm{1}}+2w_{\mathrm{c}}^{2}s_{\mathrm{0}}i_{\mathrm{0}}i_{%
\mathrm{1}}+w_{\mathrm{c}}^{4}s_{\mathrm{1}}i_{\mathrm{0}}^{2}+2w_{\mathrm{c}%
}^{2}s_{\mathrm{0}}i_{\mathrm{0}}s_{\mathrm{1}}+w_{\mathrm{c}}^{2}s_{\mathrm{%
a}}^{2}i_{\mathrm{1}}+s_{\mathrm{0}}^{2}s_{\mathrm{1}}  \notag \\
& +[w_{\mathrm{c}}^{2}i_{\mathrm{0}}+w_{\mathrm{c}}^{2}s_{\mathrm{0}}]\left[
l_{\mathrm{0}}(1-l_{\mathrm{1}})+l_{\mathrm{1}}(1-l_{\mathrm{0}})\right]
+ww_{\mathrm{c}}^{2}(1-l_{\mathrm{0}})l_{\mathrm{0}}o_{\mathrm{1}}.
\end{align}%
where $Q^{\prime }$ is obtained from $Q$\ by exchanging $0$ and $1$ subscripts and $Q$ is given by: 
\begin{align}
Q& =(1+w)[i_{\mathrm{1}}i_{\mathrm{0}}+w_{\mathrm{c}}^{2}s_{\mathrm{1}}i_{%
\mathrm{0}}+w_{\mathrm{c}}^{2}i_{\mathrm{1}}s_{\mathrm{0}}+w_{\mathrm{c}%
}^{2}s_{\mathrm{1}}s_{\mathrm{0}}+w^{2}w_{\mathrm{p}}w_{\mathrm{h}}o_{%
\mathrm{1}}o_{\mathrm{0}}]  \notag \\
& +wo_{\mathrm{1}}i_{\mathrm{1}}(1+w)+w_{\mathrm{p}}[(1-l_{\mathrm{1}%
})^{2}l_{\mathrm{0}}+wl_{\mathrm{1}}^{2}(1-l_{\mathrm{0}})].
\end{align}
The polynomials $Q$ and $Q^{\prime }$ relate $B_{m}$ with $B_{m+1}$ for even and odd levels as follows:
\begin{equation}
B_{m}\equiv B_{m+1}^{3}Q,\text{ \ (}m\text{ even), }B_{m}\equiv B_{m+1}^{3}Q,%
\text{ \ (}m\text{ odd).}  \label{amplitude_relation}
\end{equation}%
\qquad

At low temperatures, the only solutions that we have been able to find is the 2-cycle solution in which the ratios at even levels are different from those at odd levels. At high temperatures, this difference between even and odd levels vanishes so that $x_{m}(\mathrm{\alpha })=x_{m+1}(\mathrm{\alpha })=x_{m+2}(\mathrm{\alpha })$, $\alpha =$ I, O, L, R or S, and the 2-cycle solution reduces to a 1-cycle solution, as noted above.

\subsection{Free Energy}

In order to determine which phase is the stable one at any temperature, we have to find the free energy corresponding to the grand  canonical partition function \ref{PF1} of all the possible phases of the system as a function of $w$. The free energy per site, which is related to the osmotic pressure, can be easily calculated from the expressions for the total partition function $Z$ at the $(m=0)$th, $(m=1)$th and $(m=2)$th levels \cite{secondpaper}.

The total partition function of the system $Z_{0}$ can be written by considering the two $(m=0)$th branches $\mathcal{C}_{0}$ meeting at the origin, the $(m=0)$th level site. For this, we need to consider all the possible configurations in the two branches. This is done by considering all the configurations that the system can assume in the two squares that meet at the origin of the lattice. There are two configurations which corresponds to states in which a bond gets to the origin but does not cross it, each of them contributing
\begin{equation}
Z_{0}(\text{I})Z_{0}(\text{O})
\end{equation}%
to the total partition function. This term reflects the fact that such configurations look like on $\text{I}$ state as looked from one side of the origin and as an $\text{O}$ state as looked at from the opposite side. Two other configurations correspond to cases in which a bond passes through the origin and makes a bend at the origin, each contributing 
\begin{equation}
(1/w)Z_{0,\mathrm{g}}(\text{L})Z_{0,\mathrm{g}}(\text{R}),
\end{equation}%
where the factor $(1/w)$ is needed in order not to take into account the Boltzmann weight for the polymer bend at the origin twice and the subscript ``g'' refers to the gauche part of the partition function for L and R states. In fact, it is possible to separate a ``gauche'' and a ``trans'' contribution to the partial partition functions for the R and L states at any level. The ``gauche'' portion is the one corresponding to configurations such that there is a bending at the $m$th level site, while the ``trans'' portion is the one corresponding to configurations in which the two bonds coming out of the $m$th level site that we are considering are straight. It is easily seen that: 
\begin{align}
Z_{m,\text{t}}(\text{L})& =Z_{m+1}(\text{R})\{Z_{m+1}(\text{I})Z_{m+2}(\text{%
I})+w_{\mathrm{c}}^{2}Z_{m+1}(\text{S})Z_{m+2}(\text{I})  \notag \\
& +w_{\mathrm{c}}^{2}Z_{m+1}(\text{I})Z_{m+2}(\text{S})+w_{\mathrm{c}%
}^{2}Z_{m+1}(\text{S})Z_{m+2}(\text{S})+w^{2}w_{\mathrm{p}}w_{\mathrm{h}%
}Z_{m+1}(\text{O})Z_{m+2}(\text{O})\}  \notag \\
& +w_{\mathrm{p}}Z_{m+1}^{2}(\text{R})Z_{m+2}(\text{L})+Z_{m+2}(\text{R}%
)wZ_{m+1}(\text{I})Z_{m+1}(\text{O}).
\end{align}%
\begin{align}
Z_{m,\text{g}}(\text{L})& =wZ_{m+1}(\text{L})\{Z_{m+1}(\text{I})Z_{m+2}(%
\text{I})+w_{\mathrm{c}}^{2}Z_{m+1}(\text{S})Z_{m+2}(\text{I})  \notag \\
& +w_{\mathrm{c}}^{2}Z_{m+1}(\text{I})Z_{m+2}(\text{S})+w_{\mathrm{c}%
}^{2}Z_{m+1}(\text{S})Z_{m+2}(\text{S})+w^{2}w_{\mathrm{p}}w_{\mathrm{h}%
}Z_{m+1}(\text{O})Z_{m+2}(\text{O})\}  \notag \\
& +w_{\mathrm{p}}wZ_{m+1}^{2}(\text{L})Z_{m+2}(\text{R})+wZ_{m+2}(\text{L}%
)wZ_{m+1}(\text{I})Z_{m+1}(\text{O}).
\end{align}

Two configurations correspond to cases in which a bond passes through the origin but does not make any bend at the origin, one contributing 
\begin{equation}
Z_{0,\mathrm{\text{t}}}^{2}(\text{L})
\end{equation}%
and the other one contributing 
\begin{equation}
Z_{0,\mathrm{\text{t}}}^{2}(\text{R})
\end{equation}%
depending on the position of the bonds at the origin. Finally, one configuration corresponds to the presence of a solvent molecule at the origin, with a contribution: 
\begin{equation}
(1/\eta )Z_{0}(\text{S})^{2}.
\end{equation}%
Also in this case, the factor $(1/\eta )$ is necessary in order not to not to take into account the activity for the filler particle at the origin twice. It is then possible to write:  
\begin{align}
Z_{0}& =2Z_{0}(\text{I})Z_{0}(\text{O})+(2/w)Z_{0,\mathrm{\text{g}}}(\text{L}%
)Z_{0,\mathrm{\text{g}}}(\text{R})+Z_{0,\mathrm{\text{t}}}^{2}(\text{L}%
)+Z_{0,\mathrm{\text{t}}}^{2}(\text{R})+Z_{0}^{2}(\text{S})/\eta  \notag \\
& \equiv B_{0}^{2}Q_{2},
\end{align}%
where $B_{0}=Z_{0}(\text{L})+Z_{0}(\text{R}),$ and where we have introduced
the polynomial 
\begin{equation}
Q_{2}=2i_{0}o_{0}+\frac{2}{w}l_{0,\text{g}}r_{0,\text{g}}+l_{\text{t}%
,0}^{2}+r_{\text{t},0}^{2}+s_{0}^{2}/\eta ,
\end{equation}%
where $r_{0,\text{t}}$, $r_{0,\text{g}}$,$l_{0,\text{t}}$,$l_{0,\text{g}}$ are the trans and gauche portions of $r$ and $l$, respectively.

The total partition function can be used to obtain the thermodynamics of the system. It is clear that $Z_{0}$ is the total partition function of the system obtained by joining two branches $\mathcal{C}_{0}$ together at the origin. Now, let us imagine taking away from the lattice the two squares that meet at the origin. This leaves behind six different branches $C_{1}$. We connect these branches to form three smaller but identical lattices, the partition function of each of which is denoted by $Z_{1}$; the latter can be written in a form that is identical to that of equation for $Z_{0}$, except that the index 0 of each partial partition function is replaced by 1.

The difference between the free energy of the entire lattice and that of the three reduced lattices is just the free energy corresponding to a pair of squares. This pair of squares contains one complete site, that is the center of the lattice, and six additional ``half'' sites where each of the remaining $C_{1}$ branches has a base. So, these two squares correspond to a total of four lattice sites. Following Gujrati \cite{PDG-PRL-95,secondpaper} we can write the adimensional free energy per site without the conventional minus sign as 
\begin{equation}
\omega =\frac{1}{4}\ln \left( \frac{Z_{0}}{Z_{1}^{3}}\right) .
\end{equation}

It is possible to write 
\begin{equation}
Z_{1}=B_{1}^{2}Q_{2},
\end{equation}%
similar to the way we had expressed $Z_{0}=B_{0}^{2}Q_{2}$ earlier, and where $B_{m}$ and $Q_{2}$ are defined above. Since $B_{0}=B_{1}^{3}Q,$ see (\ref{amplitude_relation}), the free energy per site can be written as 
\begin{equation}
\omega =\frac{1}{2}\ln \left( \frac{Q}{Q_{2}}\right) .
\end{equation}

In order to determine which phase is the stable one at some temperature, we must find the free energy of all the possible phases of the system as a function of $w$. The usual thermodynamic potential per site, which is the osmotic pressure, can be obtained from $\omega $ through: 
\begin{equation}
F=-T\omega .
\end{equation}

\subsection{Phase Diagram}

The typical phase diagram that is obtained is shown in Figure \ref{Figure:omega for solvent+polymer}. The values of the parameters of the problem for this particular case are shown in the figure caption.

There are three phases that are captured by this calculation, a crystalline phase and two liquid like phases. The crystalline phase is captured by the 2-cycle recursion fix-point while the two liquid phases are obtained using the 1-cycle recursion fix-point. The crystalline phase (CR) is stable at low temperature. If the system is in the CR phase and it is heated up it melts into the liquid phase at the melting temperature $T_{\text{M}}$. The liquid phase that is stable above such transition is called the equilibrium liquid (EL) in order to distinguish it from a second liquid phase that is always metastable (and that we label as ML, the metastable liquid).

If we start from high temperature in the EL phase and crystallization is avoided at the melting temperature, the system finds itself in the supercooled liquid phase. This phase merges continuously into the ML phase at a temperature $T_{\text{MC}}$ that has many characteristics of the critical temperature predicted by the mode-coupling theory \cite{PDG-Corsi-PRE-03}. Using the model of polymer that we have introduced above, it turns out that that the density of filler particles is zero in both the CR and ML phases. These two phases have an ordered structure such that the filler is pushed out of the system and all the sites of the lattice are occupied by a monomer belonging to the polymer. The filler density is non zero in the EL phase. The filler density as a function of the temperature for the system whose free energy is shown in Figure \ref{Figure:omega for solvent+polymer} is shown in Figure \ref{Figure:phis for solvent+polymer}. It is very easy to observe in the figure how the filler density goes to zero as the temperature approaches $T_{\text{MC}}$. 

\subsection{Percolation}
 
It is then interesting to study the percolation process in the EL phase keeping in mind that this phase is the stable one above $T_{\text{M}}$ and it is metastable below $T_{\text{M}}$ and above $T_{\text{MC}}$. The effect of the change in the chemical potential of the filler particles on the phase diagram is minimal. The values of $T_{\text{M}}$ and $T_{\text{MC}}$ are changing very weakly with the value of the chemical potential so that we focus on one particular representative set of values for $a$, $b$ and $c$.

In order to calculate the filler percolation we proceed as follows (see Ref. \cite{firstpaper}). Following the Gujrati approach \cite{PDG-JCP3-98}, we introduce the probability $R_{m}\leq 1$ that a site occupied at the $m$th generation is connected to a finite cluster of occupied sites at higher generations. Then, $Z_{m}R_{m}$ denotes the contribution to the partial partition function $\mathcal{C}_{m}$ due to all those configurations in which the site at the $m$th generation is connected to a finite cluster in $\mathcal{C}_{m}$.

If we divide $Z_{m}R_{m}$ by $Z_{m}$, we obtain a recursion relation for $R_{m}$. At the fixed point the value of $R_{m}$ does not depend on $m$ anymore, and we denote it by $\text{R}$. For this particular problem, the equation for $\text{R}$ is:
\begin{equation}
R=\frac{w_{\text{c}}^{2}i^{3}+2Rw_{\text{c}}^{2}i^{2}s+w_{\text{c}%
}^{4}i^{2}s+3R^{2}w_{\text{c}}^{2}is^{2}+R^{3}s^{3}+2lr(w_{\text{c}%
}^{2}i+Rw_{\text{c}}^{2}s)+ww_{\text{c}}^{2}olr}{w_{\text{c}}^{2}i^{3}+2w_{%
\text{c}}^{2}i^{2}s+w_{\text{c}}^{4}i^{2}s+3w_{\text{c}%
}^{2}is^{2}+s^{3}+2lr(w_{\text{c}}^{2}i+w_{\text{c}}^{2}s)+ww_{\text{c}%
}^{2}olr}.
\end{equation}

We can then obtain the percolation probability from the value of $\text{R}$ by means of 
\begin{equation}
p=1-R^{2}
\end{equation}%
where we consider percolation to have occurred only if the central site of the lattice is connected to an infinite cluster on either or both sides of the lattice, as explained previously \cite{firstpaper,secondpaper}.

We have studied the dependence of the strength of the percolation process on three variables: the chemical potential of the filler particle, the degree of semiflexibility of the polymer and the strength of the interaction of filler and polymer. The results are presented in Figures \ref{Figure:polymer_solvent_mu_pvst} to \ref{Figure:polymer_solvent_inter_pvsphi}. In all the figures, the filled dots correspond to temperature intervals where the disordered phase in which the fillers are embedded is thermally stable while the empty symbols correspond to intervals in which it is metastable, that is between $T_{\text{M}}$ and $T_{\text{MC}}$. Figure \ref{Figure:polymer_solvent_mu_pvst} shows the behavior of the percolation probability of the filler particles as a function of temperature for different values of the chemical potential of the filler. An increase in the magnitude of the chemical potential makes the percolation less likely to occur.

Figure \ref{Figure:polymer_solvent_mu_pvst} shows that the temperature range at which the percolation of the filler particles occurs becomes narrower and narrower as $\mu$ becomes more and more negative. In particular, for values of $\mu$ larger in magnitude than $0.5$, no percolation occurs in the system at any temperature. Above some critical value of the chemical potential, the filler particles do not percolate anymore. From an energetic point of view, if the chemical potential of the filler particles becomes too large and negative it becomes highly unfavorable for the system to have filler particles in the equilibrium configuration and the density of the filler drops to values that are too low for the percolation to occur. The tailoring of the chemical nature of the filler, as a function of the polymer matrix to be used, seems to be of the uttermost importance.

Figure \ref{Figure:polymer_solvent_mu_pvst} demonstrates that the strength of the percolation process can still be very strong in the metastable phase. In the temperature intervals corresponding to the empty symbols in the figures, the system is not percolating in the ordered stable phase while it is percolating in the disordered metastable phase. We have observed something very similar in a parallel study \cite{secondpaper} when discussing the percolation in the case of a system made of particles of different sizes (but with no polymer matrix present in the system). In that case also, the metastable phase presents a stronger percolation than the ordered phase. These observations suggest how it might be useful to prepare the system in a metastable phase in order to obtain a stronger percolation.
From an experimental point of view, a metastable phase can be very similar to a stable one in many aspects. As long as the lifetime of the metastable phase is very long with respect to the observation time or the expected lifetime of an application produced using such a metastable phase, it might be useful to have a system in a metastable phase in order to increase the strength of a percolation process. We have also studied the percolation process as a function of the semiflexibility of the polymer chain. As we have argued elsewhere \cite{Rane-PDG-Macro-05} and above, the parameter $a\equiv \varepsilon _{\mathrm{p}}$ gives a measure of the stiffness of the polymer chains. The value $a=0$ corresponds to the least flexible polymer chain, and the original Flory's model. As the value of $a$ increases, the polymer chain becomes more and more flexible. This is easy to understand if we remember that increasing $a$ is equivalent to decreasing $\varepsilon $ while keeping $\varepsilon _{\mathrm{p}}$ constant.

Figure \ref{Figure:polymer_solvent_flex_pvst} shows the value of the percolation probability as a function of temperature for different values of $a$. An increase in the stiffness of the polymer matrix makes the percolation of the filler particles more and more unfavorable. In particular, while for $a<0.1$ the system is not percolating anymore at the lowest temperature at which the metastable phase exists, for larger values of this parameter the system is still percolating when it gets to $T_{\text{MC}}$. Even larger values of $a$ would just move the percolation threshold to higher and higher temperatures. The stiffness of the polymer matrix apparently opposes the percolation of the filler particles. While there is no percolation in the ordered crystalline phase, the increased stiffness makes the percolation more and more difficult to occur in the disorder phase as well, in both the stable and metastable regions. We are not aware of any experimental systematic study of this kind of dependence: one might argue that a decreased stiffness could help the percolation process by creating some preferential \textit{channels} for the filler particles to percolate. However, this would be misleading because the dependence on (repulsive) interaction $\varepsilon _{\text{c}}$ must also be taken into consideration in addition to the chemical potential. The chosen values of the two conspire in a such a way to the above result. 

Finally, we have studied the dependence of the percolation process on the strength of the interaction between polymer matrix and fillers. In the model, the parameter $c\equiv \varepsilon _{\mathrm{c}}/\varepsilon $ is associated with the excess interaction between the filler particles and the monomers belonging to the polymer chain. An increase in $c$ corresponds to an increase in the repulsion between the filler particles and the polymer matrix. The results summarized in figures \ref{Figure:polymer_solvent_inter_pvst} and \ref{Figure:polymer_solvent_inter_pvsphi} show that the percolation process becomes more favorable as the repulsion between filler and matrix increases. This is particularly clear in Figure \ref{Figure:polymer_solvent_inter_pvsphi} where it is possible to observe how a change of $c$ from $0.04$ to $0.08$ induces a change in the percolation threshold from $0.25$ to $0.245$. When the repulsion between matrix and filler is larger, it takes a lower density of filler in the system to obtain a percolating cluster. This should not come as a surprise as increasing the repulsion will eventually bring about a phase separation between fillers and the polymer matrix, in which the the solvent phase will be percolating. This kind of phase separation is not desirable for many application. The above-mentioned increase in percolation tendency is a precursor of this eventual phase separation in the model.

As the repulsion becomes weaker, the filler particles are better distributed in the matrix and they tend to form smaller and smaller clusters since they have almost no preference between sitting next to another filler particle or a polymer unit. Thus, from an experimental point of view, it seems necessary to introduce filler particles with quite a strong repulsion towards the matrix in order to increase the probability of having percolation in the system if a stronger tendency to percolation is desired.

Thus, it is clear once again the tailoring of the filler to the matrix appears to be fundamental to obtain optimal properties.

\section{Percolation of different-sized filler particles in the presence of a polymer matrix}

In order to study the effect of size and shape disparity on percolation of fillers in a polymer matrix, we have considered the addition of two different particles to the single-site fillers whose percolation has been described in the previous section. We have studied the percolation process in the following systems:

\begin{itemize}
\item single-site particles and \textit{square} particles in the presence of a polymer matrix

\item single-site particles and \textit{star} particles (occupying 5 lattice sites) in the presence of a polymer matrix
\end{itemize}

This choice allows us to draw some conclusion about the effects of the size and shape of the particles on the percolation properties. In the following, we will always refer to the single-site species as the ``solvent'' particles even when they are not the majority component.

\subsection{States of a site}

In order to solve the problem recursively, we have to define the different states a site can be in. In the case of square and single-site particles, every site of the lattice can be in one of seven states. If we look at an $m$th level site, this site can be occupied by either a monomer that belongs to a polymer chain and that can be in one of the states $\text{I}$, $\text{O}$, $\text{L}$ or $\text{R}$, introduced in the previous section, or by a solvent molecule, in the $\text{S}$ state, or by the edge of a square, either in the $\text{A}$ or in the $\text{B}$ state, according to the position of the square with respect to the site \cite{secondpaper}. The site under investigation is in an $\text{A}$ state if a vertex of the square is at the site and the square lies above the $m$th level. The site is in the $ \text{B}$ state if the vertex of the square is at the $m$th level site but the square lies below the site.

In the case of star and single-site particles, every site of the lattice can be in one of eight states. If we look at an $m$th level site, this site can be occupied by either a monomer that belongs to a polymer chain and that can be in one of the states $\text{I}$, $\text{O}$, $\text{L}$ or $\text{R}$, introduced above, or by a solvent molecule, in the $\text{S}$ state, or by a star, either in the $\text{A}$ or in the $\text{B}$ or in the $\text{C}$ state \cite{secondpaper}. As said above, we consider stars that occupy five
sites: a central one, named as the ``core'' and four endpoints that are all nearest neighbors of the core but (on the square and Husimi lattices) are not nearest neighbors of each other. The $\text{A}$, $\text{B}$ and $\text{C}$, states are defined as follows \cite{secondpaper}: the site under investigation is in an $\text{A}$ state if one end point of the star is at the site but the core of the star is above the $m$th level. The site is in the $\text{B}$ state if an end point of the star is at the $m$th level site but its core is below the site. Finally, the site is in the $\text{C}$ state if the core of the star is at the $m$th level site that we are considering.

We now consider the two cases separately.

\subsection{Squares and solvent molecules}

The total partition function of the system made of solvent molecules and square particles on the Husimi lattice is

\begin{equation}
Z_{N}=\sum \eta ^{N_{\text{S}}}\eta _{b}^{4N_{\text{sq}}}w_{\text{c}}^{N_{%
\text{c}}}w_{\text{d}}^{N_{\text{d}}}w_{\text{e}}^{N_{\text{e}}}w^{N_{%
\mathrm{g}}}w_{\mathrm{p}}^{N_{\mathrm{p}}}w_{\mathrm{h}}^{N_{\mathrm{h}}}.
\label{PF_Square}
\end{equation}%
The number of solvent molecules $N_{\text{S}}$ is controlled by the solvent activity $\eta =w^{-\mu }$, where $\mu $ is the chemical potential of the solvent particle. The number of square molecules $N_{\text{sq}}$ is controlled by the activity $\eta _{\text{b}}=w^{-\mu _{\text{b}}}$, where $\mu _{\text{b}}$ is the chemical potential associated with one side of the square particle. There are three different interactions between nearest neighbor contact pairs $N_{\text{c}},$ $N_{\text{d}},$ $N_{\text{e}}$ between solvent particles and monomers, square particles and monomers and solvent particles and square particles, respectively. The corresponding Boltzmann weights are $w_{\text{c}}=\exp \left( -\beta \varepsilon _{\text{c}}\right) =w^{\text{c}}$, $w_{\text{d}}=\exp \left( -\beta \varepsilon _{\text{d}}\right) =w^{\text{d}}$ and $w_{\text{e}}=\exp \left( -\beta \varepsilon _{\text{c}}\right) =w^{\text{e}}$, respectively, where the $\varepsilon ^{\prime }$s are the corresponding excess energies. The parameters that describe the nature of the polymer are the same ones introduced in the previous section. Since we are interested in the case in which the system is a pure polymer phase in the ground state at $T=0$, we always consider negative values of $\mu $ and $\mu _{\text{b}}$.

The procedure to obtain the free energy and the percolation probability is the same as explained in Ref. \cite{secondpaper}. The complete recursion relations are very cumbersome but straight forward to derive by following the procedure shown explicitly in the first two papers \cite{firstpaper, secondpaper}. What one needs to do is to consider all possibilities for a plaquette to determine their contribution to any partial partition function. As the problems becomes complex, the number of possibilities grow rapidly and patience is required to ensure that all terms are accounted for; several tens of them are present in some of the equations. We give the final recursion relations in the appendix for the benefit of a reader. 

In all the cases that we have studied, both in the presence of square and star particles, the qualitative features of the phase diagram have not changed. In all of the results, we always have three polymer phases: a crystalline one stable at low temperature, an equilibrium liquid phase stable at high temperature and a second metastable liquid phase that is never the stable one but that is still very important since the metastable extension of the equilibrium liquid below the melting temperature of the system, what we call the supercooled liquid phase, merges into the metastable phase at $T_{\text{MC}}$. The free energy versus temperature diagram for all these systems looks qualitatively the same as that shown in Figure \ref{Figure:omega for solvent+polymer} with minimal quantitative corrections to the values of $T_{\text{MC}}$ and $T_{\text{M}}$ so that, once again, we just focus on one representative set of parameters. The density of both solvent and square particles in both the crystalline and the metastable phase is identically zero and the percolation can only take place in the liquid phase, both in its equilibrium branch and in the supercooled extension below the melting temperature.

When we are looking at the interactions between the different particles and their percolation properties, there are two possible approaches:

\begin{enumerate}
\item We can consider the solvent particles and the square particles as different species with a nonzero excess energy and consider the separate percolation of the two kinds of particles in the presence of the other ones.

\item We can consider the solvent and square particles to be of the same chemical nature and consider the total percolation probability when particles of different sizes and same chemical nature are present in the system. Here the cluster is formed by both particles. One can also study the percolation of each kind of particles separately.
\end{enumerate}

The first question that can be asked is the following: does a system containing just square particles exhibit any percolation? In order to study this we have set the activity for the solvent molecules equal to zero and studied the system of square particles embedded in a polymer matrix.

The results of this calculation are the following: regardless of the value of their chemical potential, the square particles alone never percolate at any temperature even if their chemical potential becomes extremely small in magnitude. We think this might be related to the nature of the polymer we have considered here and of course is also related to our choice to always consider negative values of $\mu $ and $\mu _{\text{b}}$ in order to have a pure polymer phase in the ground state at $T=0$. Having an infinite molecular weight polymer chain makes it almost impossible for these \textit{bulky} particles to form an infinitely large percolating network. By considering a system of polymer chains with a finite degree of polymerization, we might have percolation of these larger particles as well. One might also think about considering a less restrictive condition of connectivity for percolation. Here, as well as in any other calculation carried on during this research \cite{firstpaper, secondpaper}, we have considered two particles to be part of a cluster only if they are nearest neighbors of each other on the lattice. It is possible to define the condition for percolation in less strict terms: for example it is possible to consider percolation of second nearest neighbors or even of particles farther apart with respect to each other on the lattice. However, we will not do that here.

The first problem to study is the problem of the percolation of the complete system in which we consider the solvent particles and the square particles to be of the same chemical nature and just different in size. To some extent, we can think about this system as a system of particles with a bimodal distribution of sizes. In order to consider this problem, we set the excess energy for the solvent-square interaction equal to zero and we consider, out of all the possible configurations of the system at the $m$th level, those configurations corresponding to a percolating cluster of solvent and square particles.

Figure \ref{Figure:squares+solvent_total_temperature} shows the percolation probability for the complete system as a function of the temperature. The values of the parameters of the system are shown in the figure caption. One immediately notices how the temperature interval in which the system is percolating is much wider than in the case of the percolation of the single-site filler particles studied in the previous paper \cite{secondpaper}. If we consider the dependence of the percolation probability on the total
density of filler, the corresponding results are shown in Figure \ref{Figure:squares+solvent_total_density}.

The value of the percolation threshold is much lower than in the case of the single-site fillers alone, for the same value of the chemical potential (per site) for the particles. Thus, the results show that size disparity lowers the percolation threshold and makes percolation more likely to occur, even in the presence of a polymer matrix.

It is also possible to study the effect of the presence of the larger particles on the percolation of the smaller ones as compared to the case in which only the smaller particles are present in the system.

Figures \ref{Figure:Squares+solvent_solvent_Tcomparison} and \ref{Figure:Squares+solvent_solvent_phicomparison} show the percolation probability for the smaller filler particles as a function of temperature and filler density, respectively, for three different cases: in one case, only the smaller filler particles are present; in the second case both particles represent but they do not have any interaction; and in the third case both the particles are present and they have a repulsive interaction between themselves.

It is clear from the results how the presence of the larger particles makes it easier for the smaller particles to percolate, lowering their percolation threshold. The effect of the interactions between the different particles seems weaker since we do not see a significant difference between the curves corresponding to the cases in which the two kinds of particles interact and do not interact between themselves. Even if the effect is small, though, it is clear how the presence of repulsive interaction between the particles slightly enhances the percolation process, as seen in Figure \ref{Figure:Squares+solvent_solvent_phicomparison} where the percolation threshold is slightly lower in the case in which repulsive interactions are present between the particles. Figure \ref{Figure:squares+solvent_comparison} shows how much the percolation probability is increased by the presence of the larger particles. The filler density is, in both cases, the total density associated with all the filler particles present in the system.

\subsection{Stars and solvent molecules}

For a system made of solvent particles, star particles and polymer, the total partition function is identical to that for squares \ref{PF_Square}, except that $N_{\text{sq}}$ is replaced by the number of star molecules $N_{\text{st}}$ This number is controlled by the activity $\eta _{\text{b}}=w^{-\mu _{\text{b}}}$, where $\mu _{\text{b}}$ is the chemical potential associated with an arm, which also happens to be a bond, of the star particle. All other terms in \ref{PF_Square} have the same meaning as above, except that stars replace squares. Since we are interested, as before, in the case in which the system is a pure polymer in the ground state at $T=0$,
we always consider negative values of $\mu $\ and $\mu _{\text{b}}$.

We give the final recursion relations in the appendix for the benefit of the reader. 

As in the case of the problem of square particles, the first question that we have asked is the following: does a system made of just star particles in the presence of a polymer matrix exhibit any percolation? In order to study this problem, once again we have set the activity for the solvent molecules equal to zero and studied the system of star particles embedded in a polymer matrix.

Exactly like in the case of square particles, we have obtained that regardless of the value of their chemical potential, the star particles alone never percolate at any temperature. As explained in the previous subsection, we think this might be related to the nature of the polymer we have considered here.

As in the previous case, we have studied the problem of the percolation of the complete system in which we consider the solvent particles and the star particles to be of the same chemical nature and just different in size. In order to study this problem, we set the excess energy for the solvent-star interaction equal to zero and we consider, out of all the possible configurations of the system at the $m$th level, those configurations corresponding to a percolating cluster of solvent and star particles. Figure \ref{Figure:stars+solvent_total_temperature} shows the percolation probability for the complete system of solvent, stars, and polymer, as a function of the temperature. The values of the parameters for this system are shown in the figure caption. We immediately notice how the temperature interval in which the system is percolating is much wider than in the case of the percolation of the single-site filler particles but at the same time much narrower than that of the system of squares and solvent molecules studied in the previous section and shown in Figure \ref{Figure:squares+solvent_total_temperature}. The results for the dependence of the percolation probability on the total density of filler are shown in Figure \ref{Figure:stars+solvent_total_density}. The value of the percolation threshold is much lower than in the case of the single-site fillers alone, for the same value of the chemical potential (per site) for the particles. Thus, the results show how the presence of size disparity, in this system as in the system containing the square particles, lowers the percolation threshold and makes percolation more likely to occur.

It is also possible to study the effect of the presence of the larger particles on the percolation of the smaller ones as compared to the case in which only the smaller particles are present in the system, as done before for the case of the square particles. Figures \ref{Figure:stars+solvent_solvent_tcomparison} and \ref{Figure:stars+solvent_solvent_phicomparison} show the percolation probability for the smaller filler particles as a function of temperature and filler density, respectively, for three different cases: in one case only the smaller filler particles are present, in the second case both particles are present but they do not have any interaction, and in the third case both the particles are present and they have a repulsive interaction between themselves.

A careful observation of Figure \ref{Figure:stars+solvent_solvent_phicomparison} and its comparison with Figure \ref{Figure:Squares+solvent_solvent_phicomparison} shows a very interesting result: while the addition of the square particles has a minimal effect on the percolation properties of the smaller particles and induces a slight decrease in the percolation threshold for the smaller particles, the addition of star particles increases the percolation threshold for the smaller particles making percolation more unlikely to occur. This difference may be related to the fact that these two particle shapes that have been used, squares and stars, have a different size but the same number of nearest neighbors interaction that they can create with other species, eight in both cases on a Husimi lattice. It is clear form the results how the presence of the larger particles makes it easier for the smaller particles to percolate, lowering their percolation threshold. As we have done before, we have to point out that at the present stage it is more proper to state that this might not be the only important factor affecting the percolation threshold and that a host of properties might be involved. The effect of the interactions between the different particles seems not very important since we do not see a significant difference between the curves corresponding to the cases in which the two sets of particles interact and do not interact between themselves. Even if the effect is small, though, it is clear how the presence of repulsive interactions between the particles slightly enhances the percolation process, as seen clearly in Figure \ref{Figure:stars+solvent_solvent_phicomparison} where the percolation threshold is slightly lower in the case in which repulsive interactions are present between the particles. Figure \ref{Figure:stars+solvent_comparison} shows how much the percolation probability is increased by the presence of the larger particles. The filler density in both cases is the total density associated with both the solvent particles and the larger ones (either squares or stars).

It is very useful to compare the percolation probability for the two composite systems that we have described in these two last sections. Figure \ref{Figure:squaresvsstars_phi} shows the percolation probability as a function of the total filler density for the two systems that we have investigated. It is clear how the percolation threshold is lower in the case of the system containing the square particles as compared to the case of the system containing the star particles. It is possible to argue that the star particles are \textit{larger} than the square particles since the former occupy five sites of the lattice (the core and the four end-points) while the square particles occupy only four sites. As observed in experiments, an increase in polydispersity and in the size disparity of the filler particles is usually associated with a decrease in the percolation threshold and our model calculation is able to capture this feature.

\section{Conclusions}

We have developed a lattice approach \cite{firstpaper, secondpaper} that allows us to study the effects of size and shape of particles and their energetics on the percolation properties of a composite system.

The approaches that have been used so far in the literature have a wide range of limitations that range from the use of random mixing approximation to the need to have incompressible systems to the requirement for the monomers and voids to have the same size. The most important limitation of all existing theories is represented by the necessity to go through some averaging process on scales that are much larger than the size of nanoscopic fillers found in many of the most advanced composites that are currently under investigation. Molecular dynamics and Monte-Carlo simulations can be used in order to obtain a better understanding of the thermodynamics of the problem; however, both these techniques are extremely time consuming and require finite size scaling that is not obvious in the case of complex systems. Also, simulations usually require an unreasonably large amount of free volume and cannot be carried out in the nearly incompressible limit.

Our approach allows us to describe how percolation is affected by the presence of size and shape disparity between the species that are present in a system. This approach is very powerful since the calculations are straightforward and exact. The solution that we obtain represents an approximate theory on the original lattice. This kind of calculation represents a new method to deal with \emph{nanoscales} where all the interactions can be taken into account and where the \emph{nanoscopic} details of composite systems can be easily taken into account.

Our results are able to reproduce most of the experimental findings that have been observed in the literature. In particular:

\begin{itemize}
\item the percolation threshold decreases as the size disparity between different particles present in the system increases. We have proved this by studying systems made of single-site, square shaped or star shaped filler particles;

\item the percolation threshold decreases with the aspect ratio of the particles present in the system. We have obtained this results through the comparison of the behavior of systems containing either the star or the square particles;

\item the stiffness of the polymer matrix as well as the nature of the interactions between fillers and matrix affect the percolation properties of the system;

\item in many different systems it is possible to observe the presence of a metastable phase that is percolating very strongly while the stable phase has a very weak percolating cluster (a much lower value of $p$) or even no percolating cluster at all. This observation suggests that it might be useful to prepare a system in a metastable state in some cases in which a percolating network of particles is needed in order to enhance one or more physical properties of the composite system.
\end{itemize}

There are still numerous aspects of this problem that are worth investigating. The effects produced by branching and finite loops have not been investigated so far to the best of our knowledge. One aspect that will be very interesting to study is the effect of the finite chain length of the polymer matrix on the percolation of the filler particles. When the polymer chains are short, either linear or branched, they can give rise to gelation, the formation of an infinite network of polymer chains. The gelation of the polymer chains and the percolation of the filler particles are two independent processes but it is easy to understand how these two processes
might strongly affect one another. Our polymer model has been recently implemented in order to describe the presence of finite polymer chains \cite{PDG-Rane-Corsi-PRE-03}. The current model describes a polydisperse polymer system with an equilibrium degree of polymerization but the model can be easily extended in order to be able to describe a monodisperse polymer system. In the case of the infinitely long polymers studied here, the larger particles never percolate in the absence of the small ones when they are inserted in the polymer matrix. We feel that this could be related to the nature of the polymer chains used and that the inclusion of finite length polymers might in principle produce new and exciting results. Another natural extension of this model will involve the study of the percolation process in the presence of even larger particles even though it has to be kept in mind that an increase in the size of the particles will lead to an increase of the complications to be faced in the description of the system.

\newpage

\bibliographystyle{unsrt}
\bibliography{bio}

\newpage

\begin{figure}
FIGURE CAPTIONS
\caption{Portion of an infinite lattice known as Husimi tree.\hfill }
\label{Husimi}
\caption{Possible configurations of the site at the $m$th level of the tree for a system  containing polymer chains and monomeric solvent molecules: $\text{I}$ state, $\text{O}$ state, $\text{L}$ state, $\text{R}$ state and $\text{S}$ state.}
\label{Figure: Polymer+solvent states}
\caption{Free energy as a function of temperature for the polymer system with monomeric filler particles. The curves for the crystalline phase, the equilibrium liquid and the metastable liquid phases are shown for $a=0.20,\mu =-0.50,b=0.07$.}
\label{Figure:omega for solvent+polymer}
\caption{Solvent density as a function of temperature for the polymer system with monomeric filler particles. The curves for the crystalline phase and the equilibrium liquid
are shown for $a=0.20,\mu=-0.50,b=0.07$.} 
\label{Figure:phis for solvent+polymer}
\caption{Dependence of the percolation probability for monomeric filler particles in the presence of a polymer matrix on the temperature of the system for different values of the chemical potential of the filler particles. For all the curves: $a=0.1$, $b=0$, $c=0.07$.}
\label{Figure:polymer_solvent_mu_pvst}
\caption{Dependence of the percolation probability for monomeric filler particles  in the presence of a polymer matrix on the temperature of the system for different values of the flexibility of the polymer chains. For all the curves: $\mu=-0.2$, $b=0$, $c=0.07$.}
\label{Figure:polymer_solvent_flex_pvst}
\caption{Dependence of the percolation probability for monomeric filler particles in the presence of a polymer matrix on the temperature of the system for different values of the interaction energy between filler particles and polymer matrix. For all the curves: $\mu=-0.2,a=0.2,b=0$.}
\label{Figure:polymer_solvent_inter_pvst}
\caption{Dependence of the percolation probability for monomeric filler particles in the presence of a polymer matrix on the volume fraction of filler for different values of the interaction energy between filler particles and polymer matrix. For all the curves: $\mu=-0.2,a=0.2,b=0$.}
\label{Figure:polymer_solvent_inter_pvsphi}
\caption{Dependence of the percolation probability on the temperature for the system containing monomeric filler particles along with square particles. Here $\mu=-0.2,\mu_{\text{b}}=-0.2,a=0.2,b=0,c=d=0.07,e=0$.}
\label{Figure:squares+solvent_total_temperature}
\end{figure}                                                                                                               

\begin{figure} 

\caption{Dependence of the percolation probability on the total density of filler for the system containing monomeric filler particles along with square particles. Here $\mu=-0.2,\mu_{\text{b}}=-0.2,a=0.2,b=0,c=d=0.07,e=0$.}
\label{Figure:squares+solvent_total_density}
\caption{Effect of the addition of a second filler species and of the interactions between fillers on the percolation probability of a composite system as a function of temperature: squares and solvent.}
\label{Figure:Squares+solvent_solvent_Tcomparison}
\caption{Effect of the addition of a second filler species and of the interactions between fillers on the percolation probability of a composite system as a function of the filler density: squares and solvent.}
\label{Figure:Squares+solvent_solvent_phicomparison}
\caption{Percolation probability as a function of the filler density for the system containing monomeric fillers and square particles compared with the percolation probability of the system containing only the smaller particles.}
\label{Figure:squares+solvent_comparison}
\caption{Dependence of the percolation probability on the temperature for the system containing monomeric filler particles along with star particles. Here $\mu=-0.2,\mu_{\text{b}}=-0.2,a=0.2,b=0,c=d=0.07,e=0$.}
\label{Figure:stars+solvent_total_temperature}
\caption{Dependence of the percolation probability on the total density of filler for the system containing monomeric filler particles along with star particles. Here $\mu=-0.2,\mu_{\text{b}}=-0.2,a=0.2,b=0,c=d=0.07,e=0$.}
\label{Figure:stars+solvent_total_density}
\caption{Effect of the addition of a second filler species and of the interactions between fillers on the percolation probability of a composite system as a function of temperature: stars and solvent.}
\label{Figure:stars+solvent_solvent_tcomparison}
\caption{Effect of the addition of a second filler species and of the interactions between fillers on the percolation probability of a composite system as a function of the filler density: stars and solvent.}
\label{Figure:stars+solvent_solvent_phicomparison}
\caption{Percolation probability as a function of the filler density for the system containing monomeric fillers and star particles compared with the percolation probability of the system containing only the smaller particles.}
\label{Figure:stars+solvent_comparison}
\caption{Percolation in systems made of particles of different shape: effect of the difference in particles sizes and shapes on the percolation probability of the system as a function of the filler density.}
\label{Figure:squaresvsstars_phi}

\end{figure}

\begin{figure}[tbp]
\begin{center}
\includegraphics[width=4in]{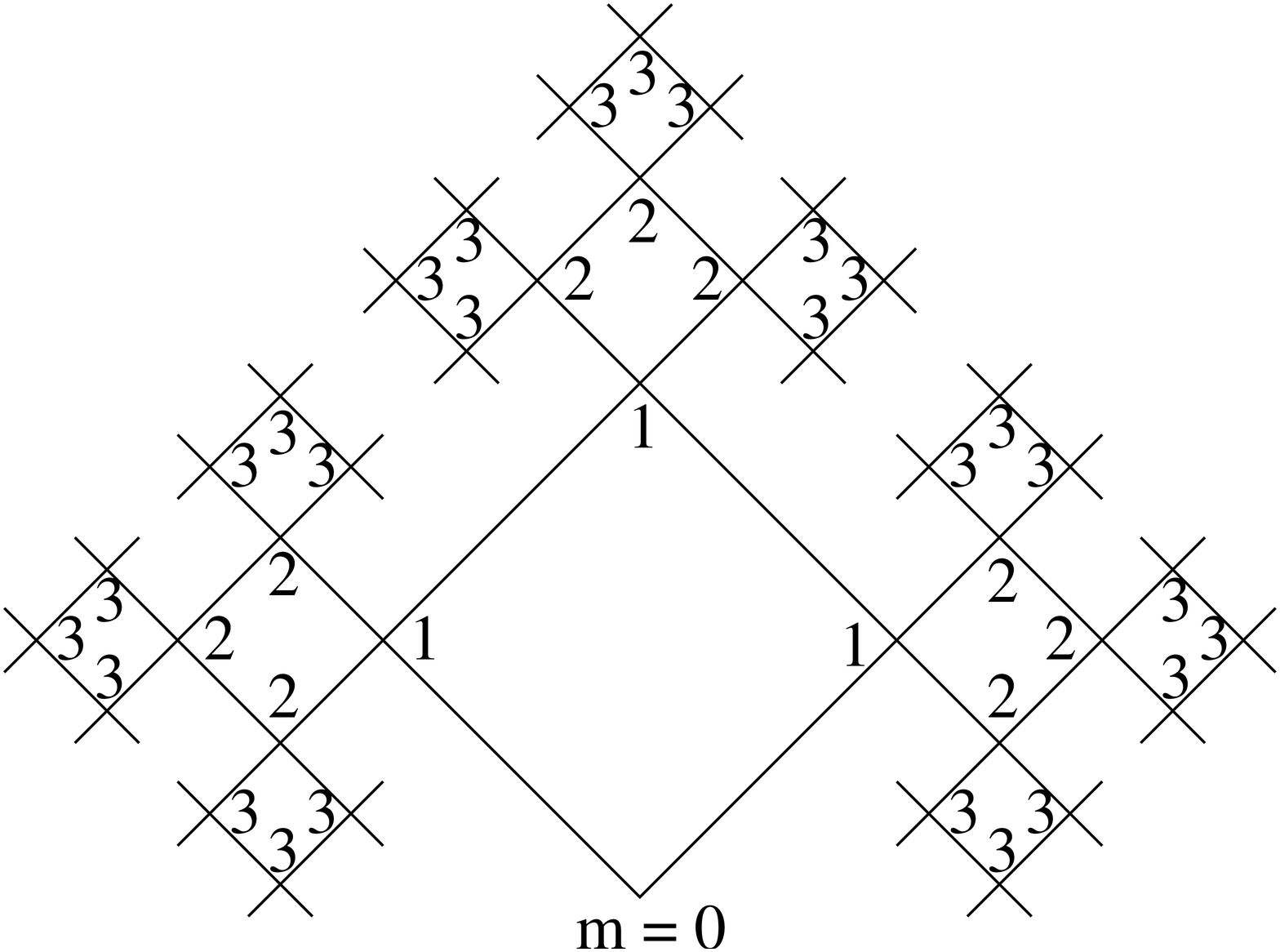}
\end{center}
\par
FIG. 1 \vspace{5cm}
\end{figure}

\newpage

\begin{figure}[tbp]
\begin{center}
\includegraphics[width=8.5cm]{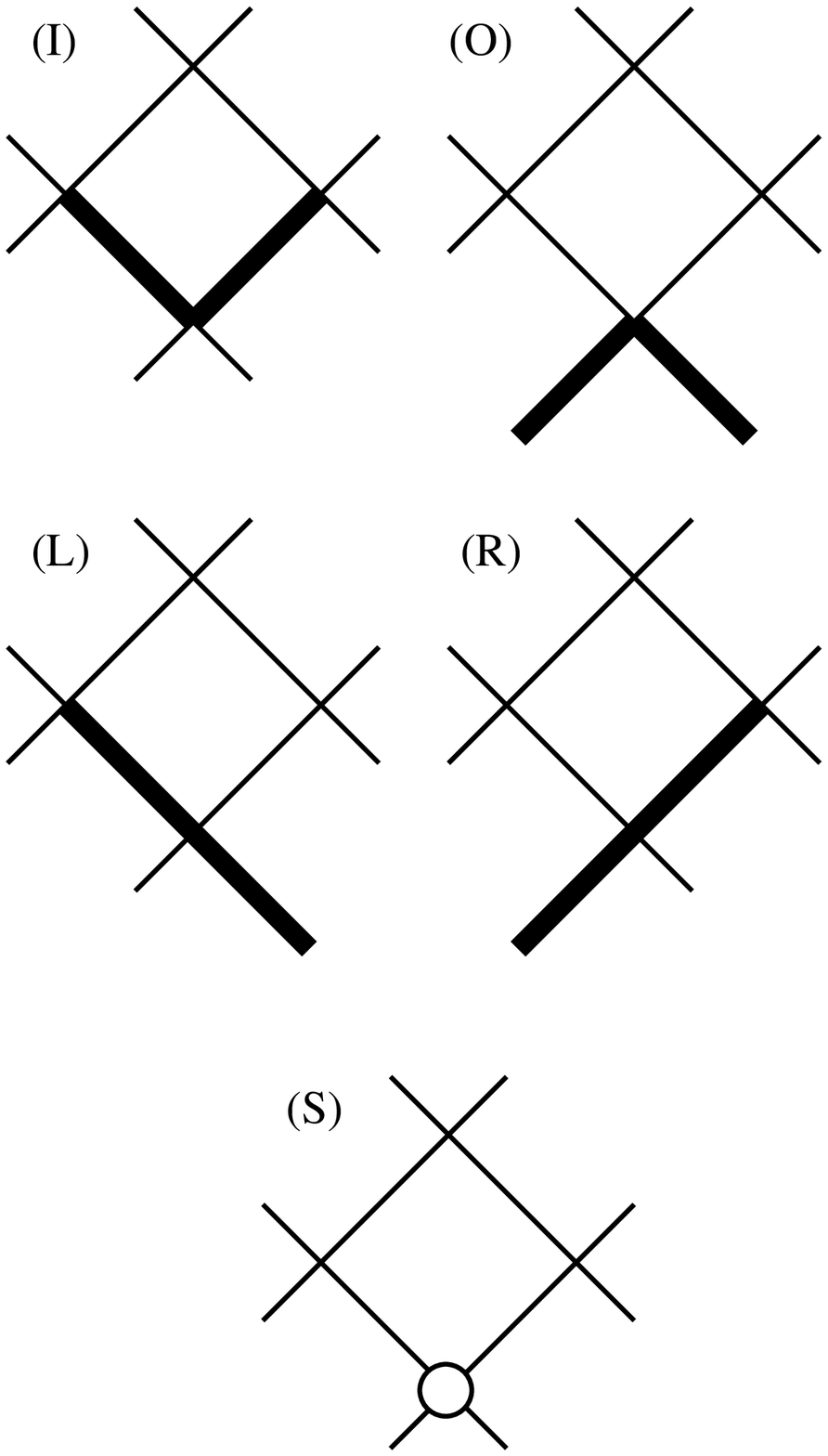}
\end{center}
\par
FIG. 2 \vspace{5cm}
\end{figure}

\newpage

\begin{figure}[tbp]
\begin{center}
\includegraphics[width=8.5cm,angle=270]{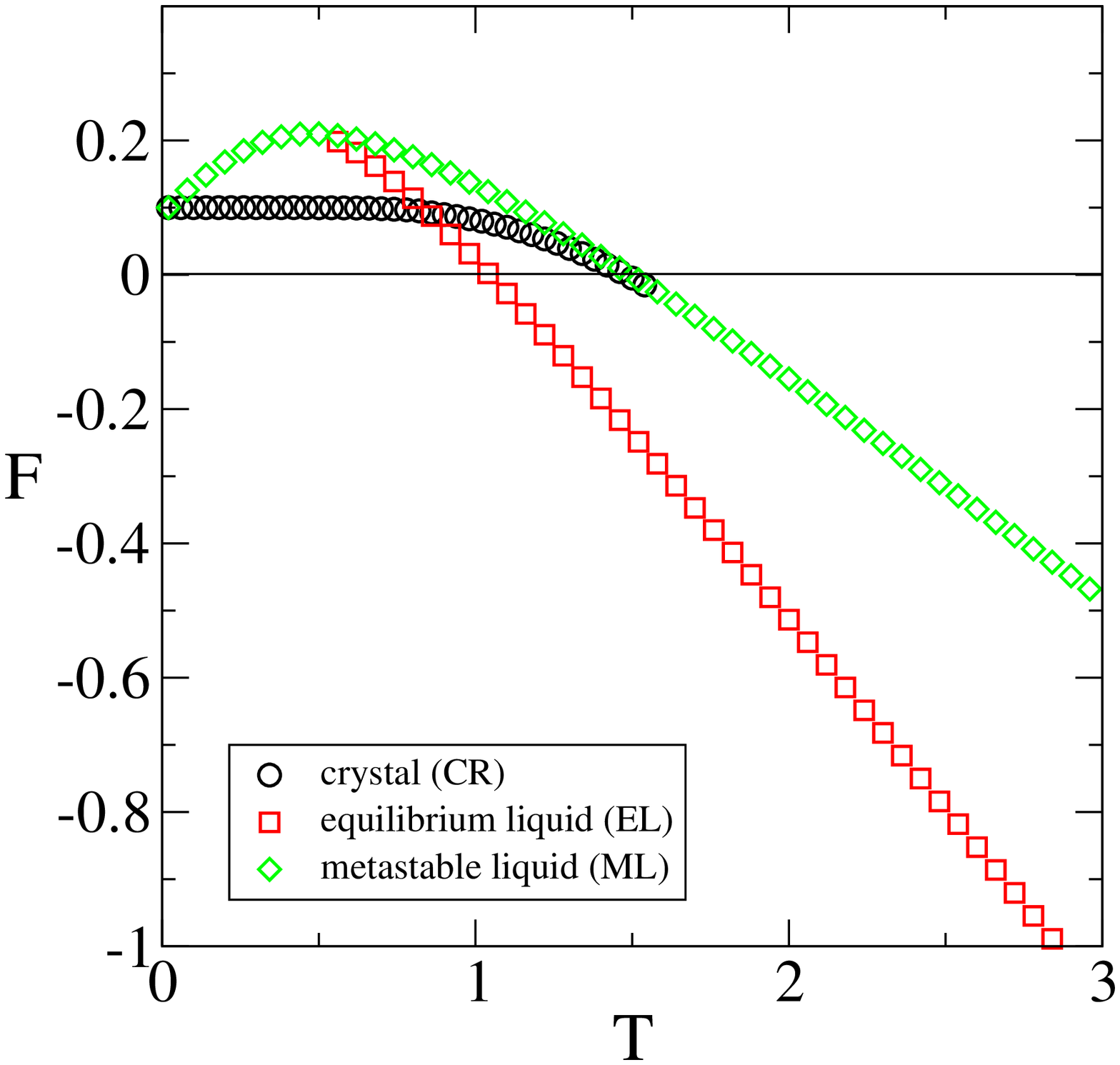}
\end{center}
\par
FIG. 3 \vspace{5cm}
\end{figure}

\newpage

\begin{figure}[tbp]
\begin{center}
\includegraphics[width=8.5cm,angle=270]{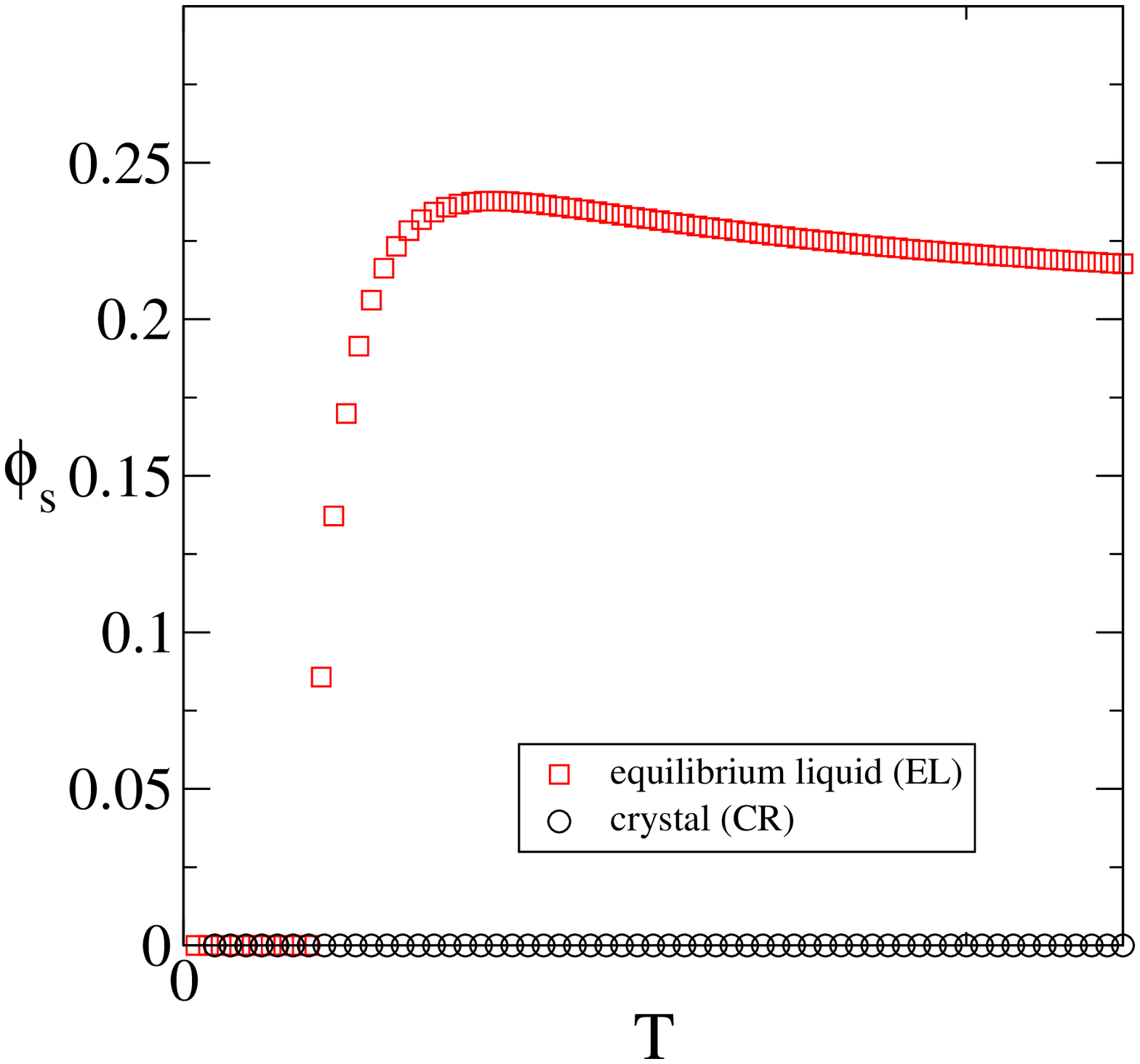}
\end{center}
\par
FIG. 4 \vspace{5cm}
\end{figure}

\newpage

\begin{figure}[tbp]
\begin{center}
\includegraphics[width=8.5cm,angle=270]{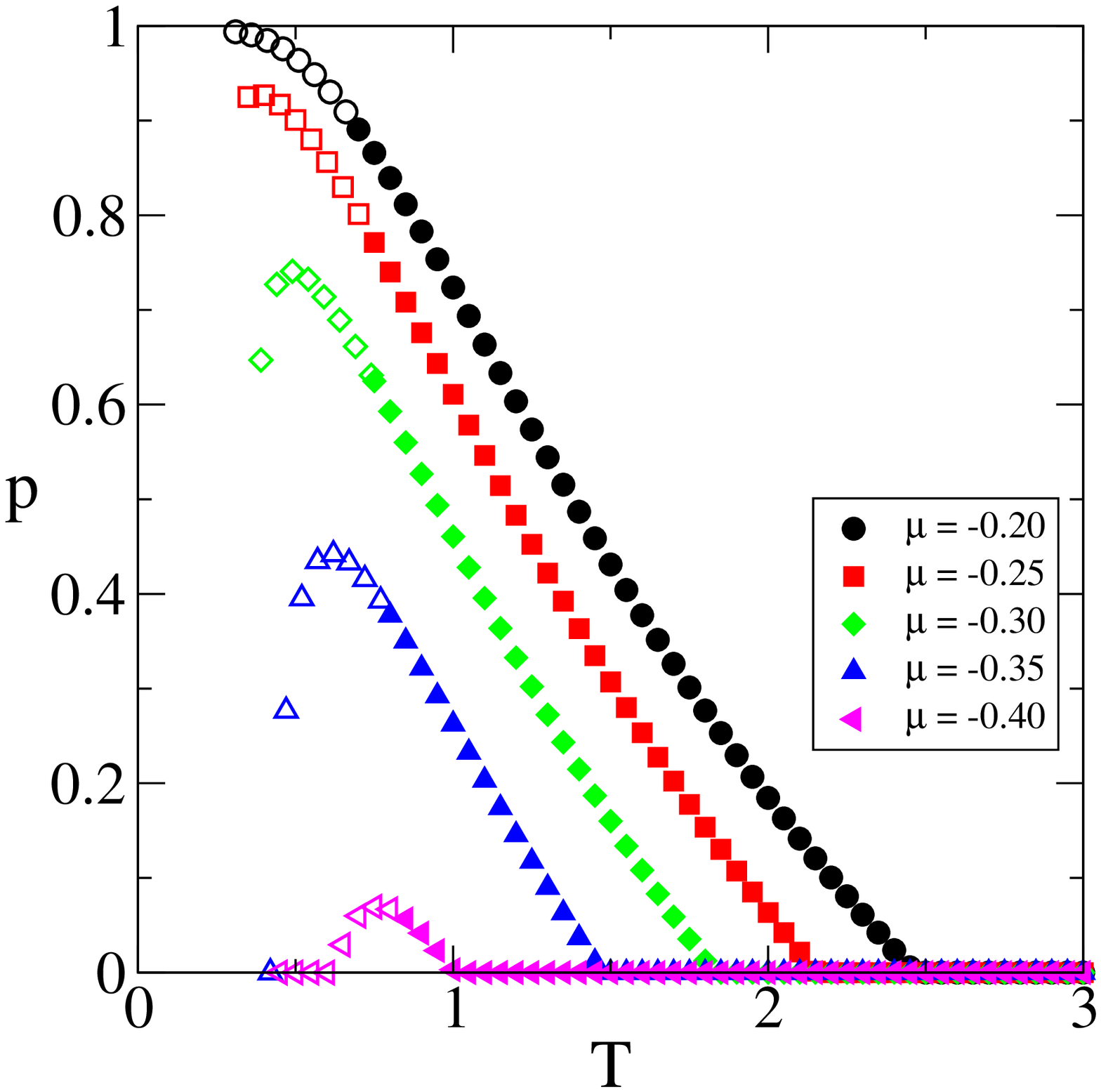}
\end{center}
\par
FIG. 5 \vspace{5cm}
\end{figure}

\newpage

\begin{figure}[tbp]
\begin{center}
\includegraphics[width=8.5cm,angle=270]{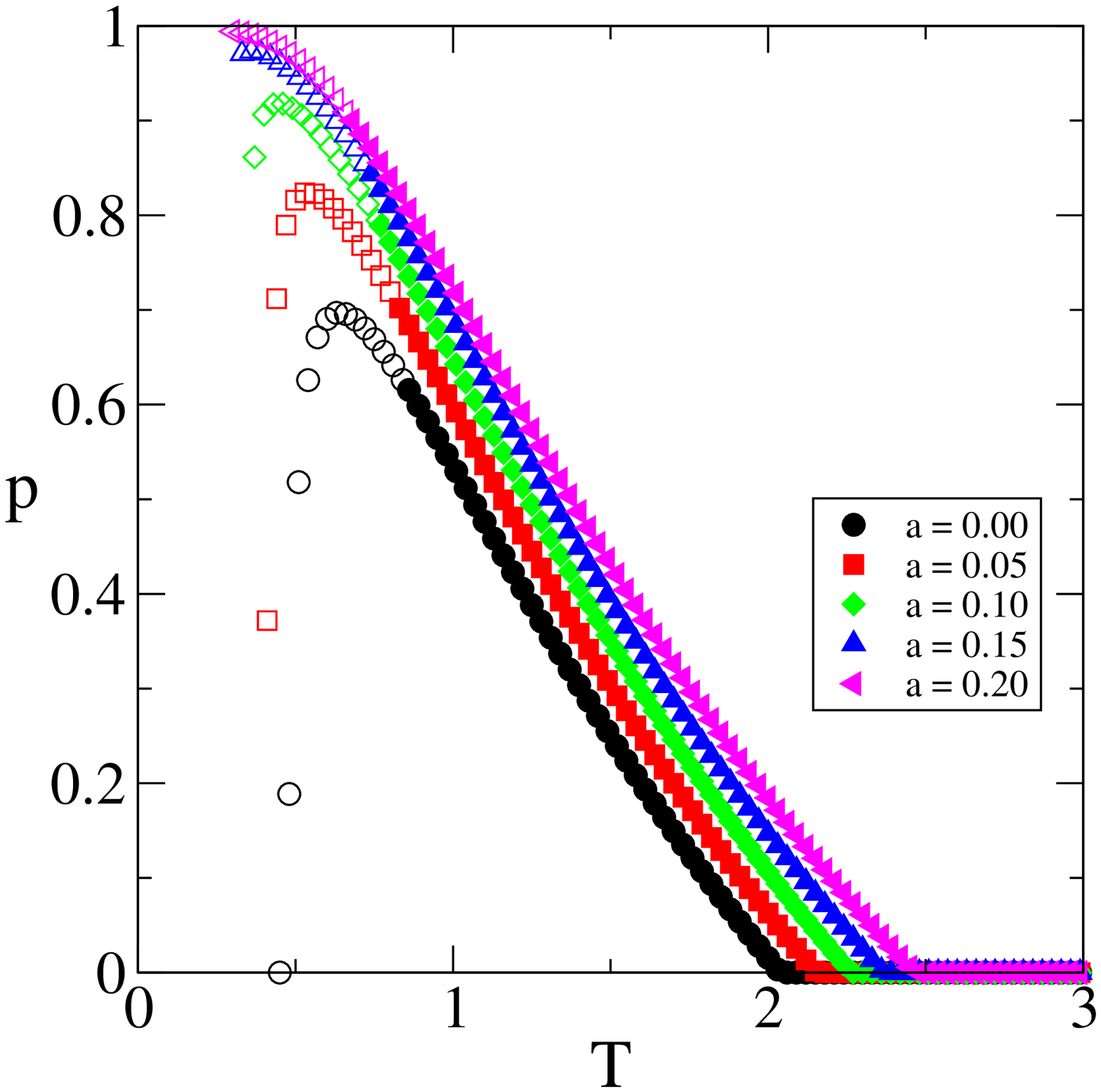}
\end{center}
\par
FIG. 6 \vspace{5cm}
\end{figure}

\newpage

\begin{figure}[tbp]
\begin{center}
\includegraphics[width=8.5cm,angle=270]{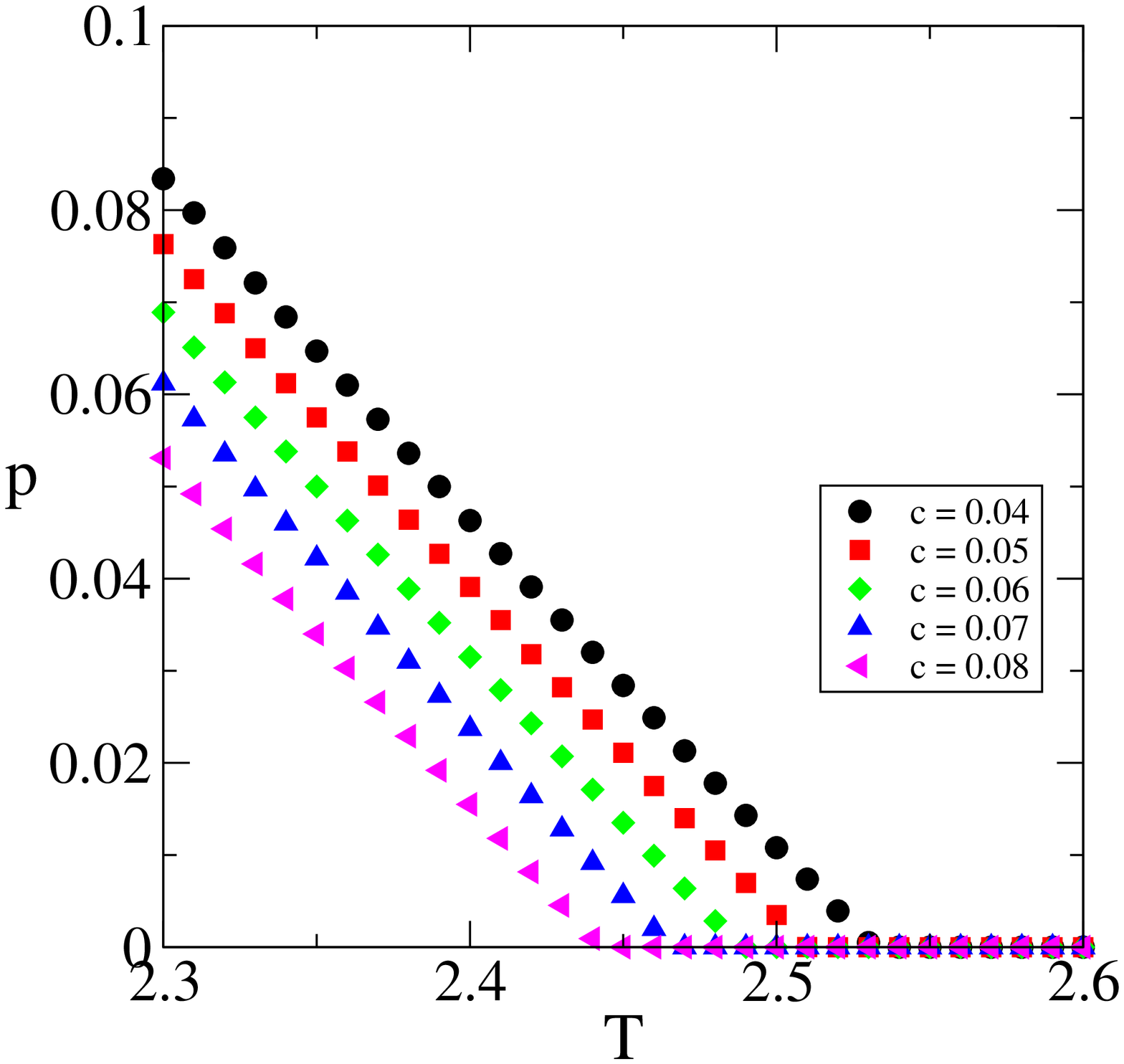}
\end{center}
\par
FIG. 7 \vspace{5cm}
\end{figure}

\newpage

\begin{figure}[tbp]
\begin{center}
\includegraphics[width=8.5cm,angle=270]{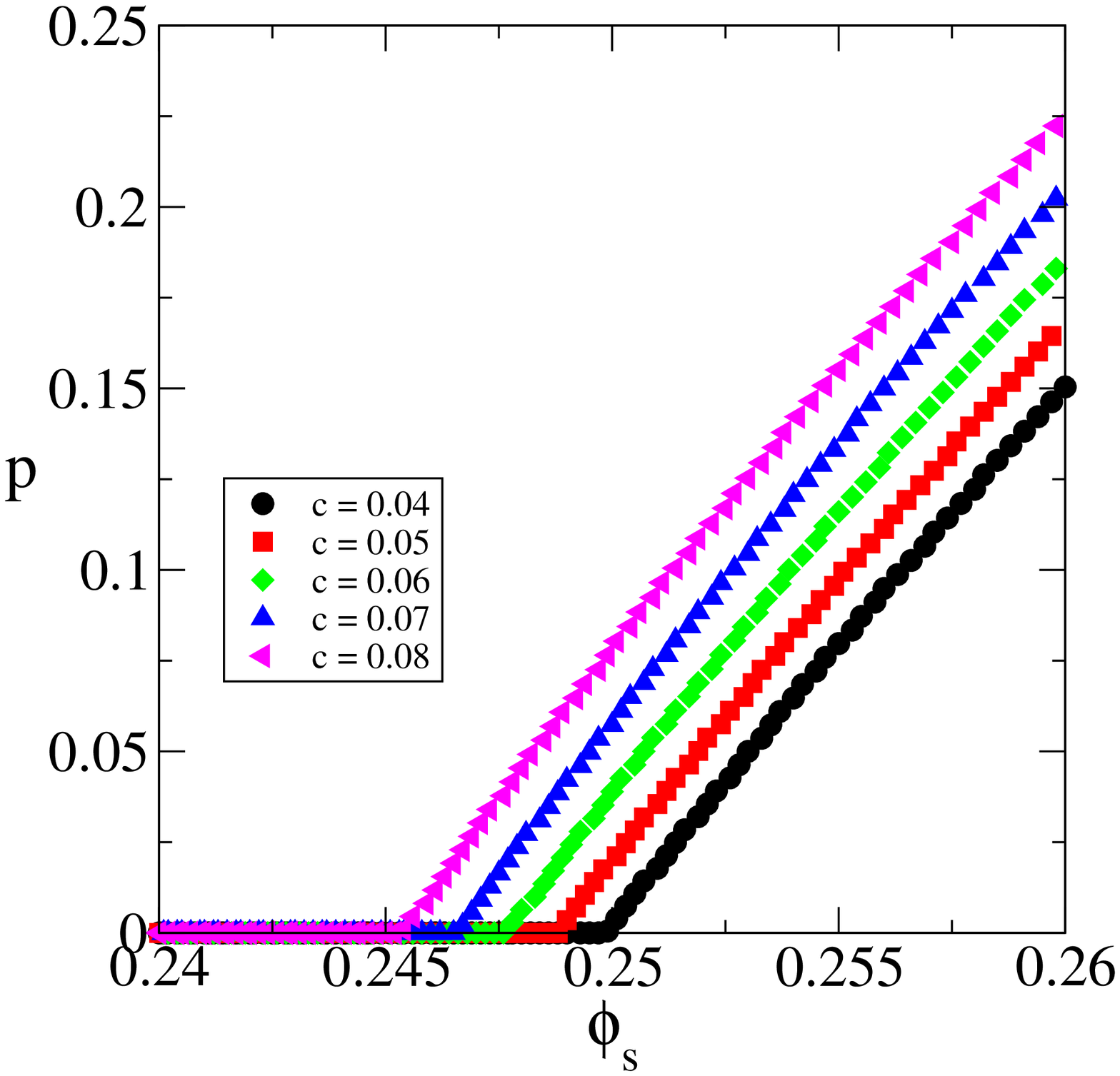}
\end{center}
\par
FIG. 8 \vspace{5cm}
\end{figure}

\newpage

\begin{figure}[tbp]
\begin{center}
\includegraphics[width=8.5cm,angle=270]{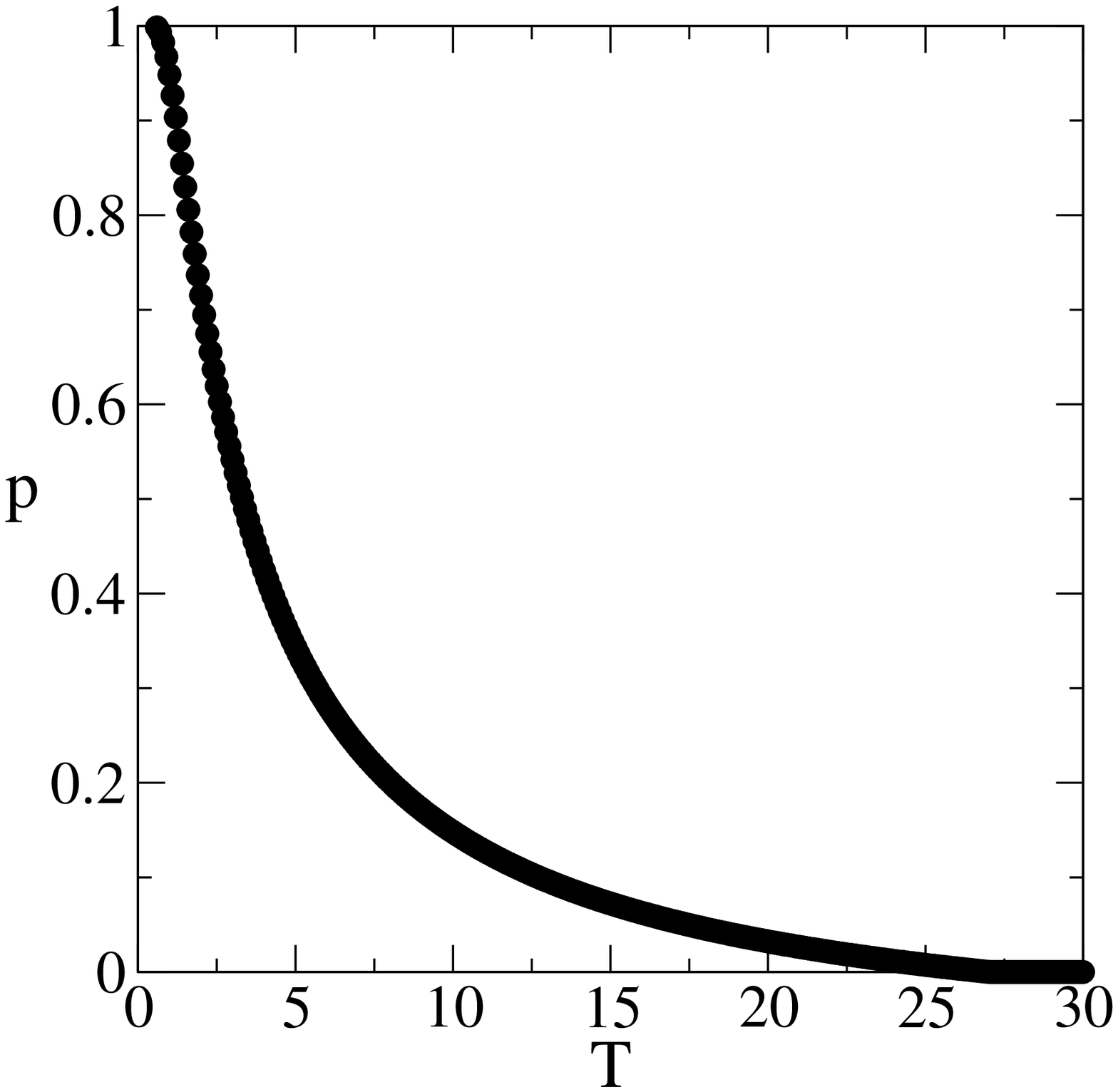}
\end{center}
\par
FIG. 9 \vspace{5cm}
\end{figure}

\newpage

\begin{figure}[tbp]
\begin{center}
\includegraphics[width=8.5cm,angle=270]{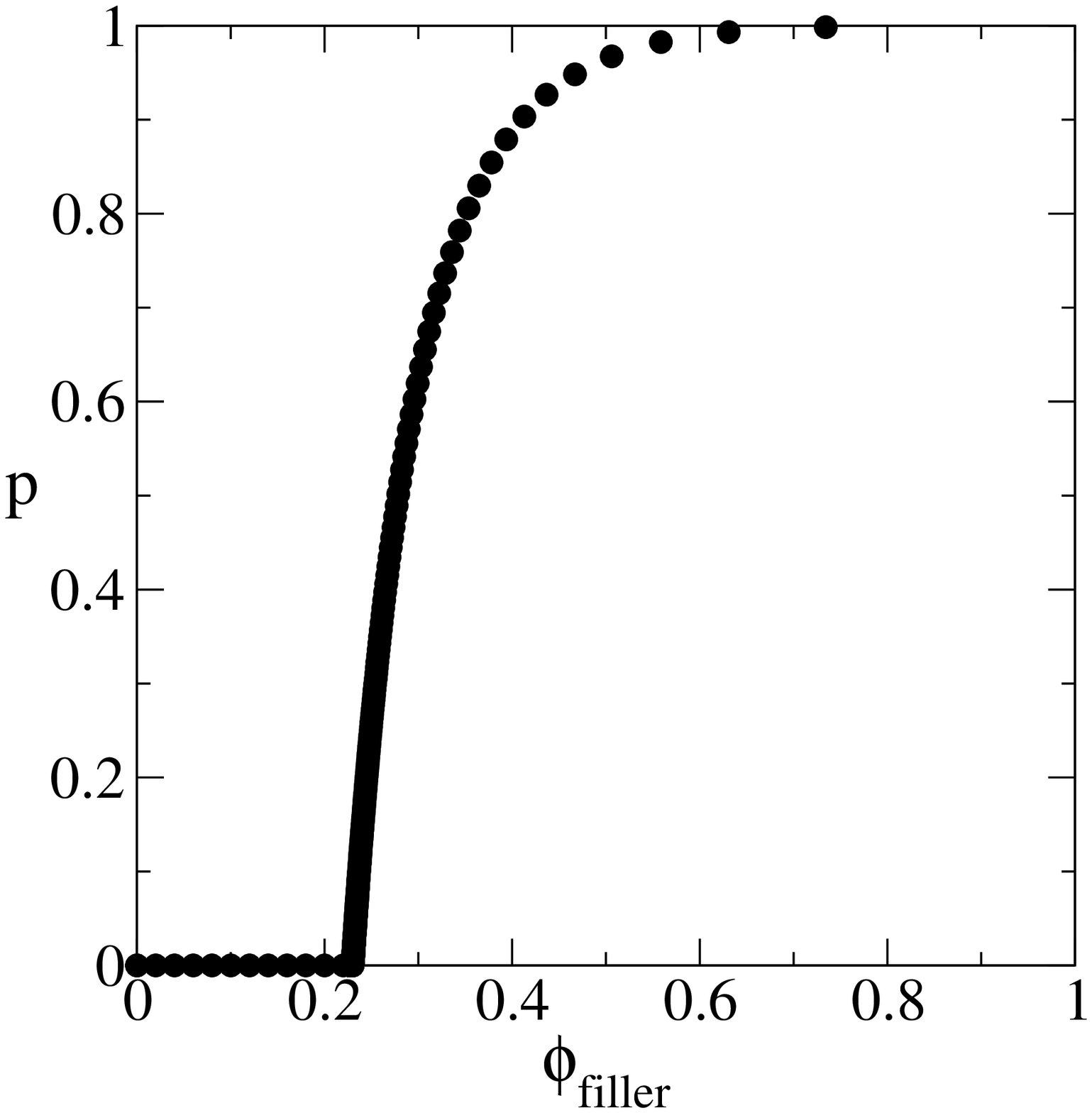}
\end{center}
\par
FIG. 10 \vspace{5cm}
\end{figure}

\newpage

\begin{figure}[tbp]
\begin{center}
\includegraphics[width=8.5cm,angle=270]{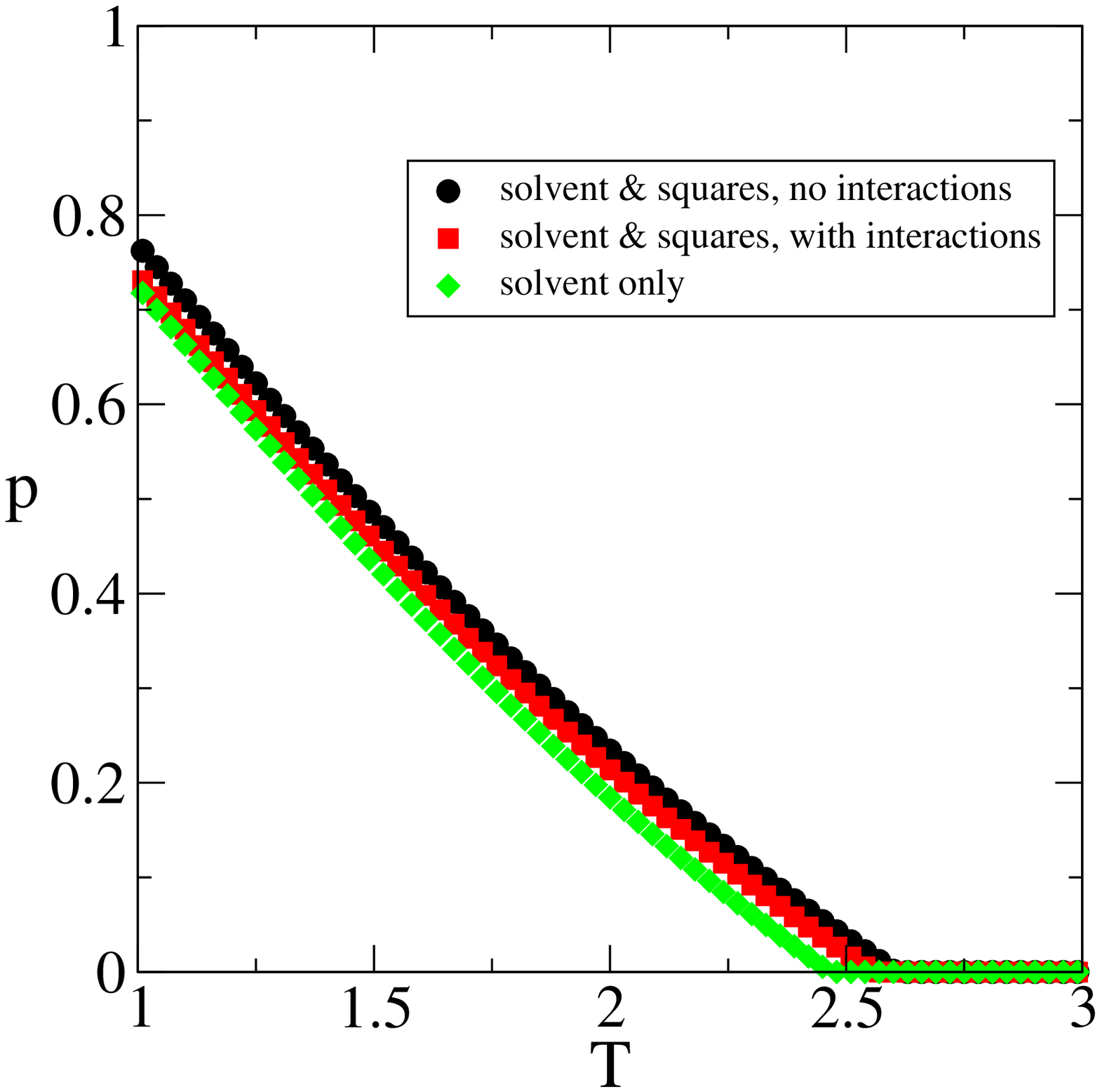}
\end{center}
\par
FIG. 11 \vspace{5cm}
\end{figure}

\newpage

\begin{figure}[tbp]
\begin{center}
\includegraphics[width=8.5cm,angle=270]{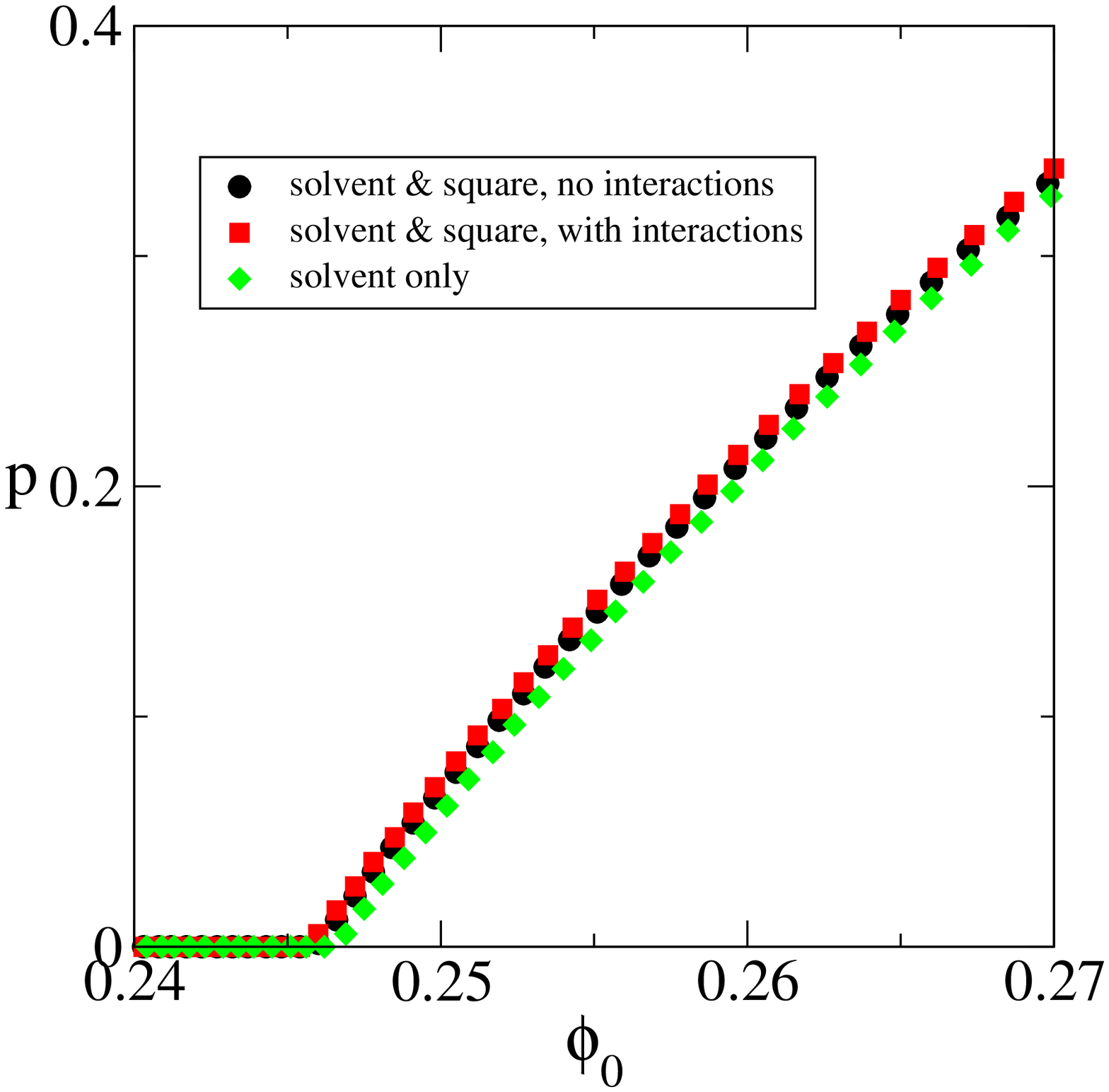}
\end{center}
\par
FIG. 12 \vspace{5cm}
\end{figure}

\newpage

\begin{figure}[tbp]
\begin{center}
\includegraphics[width=8.5cm,angle=270]{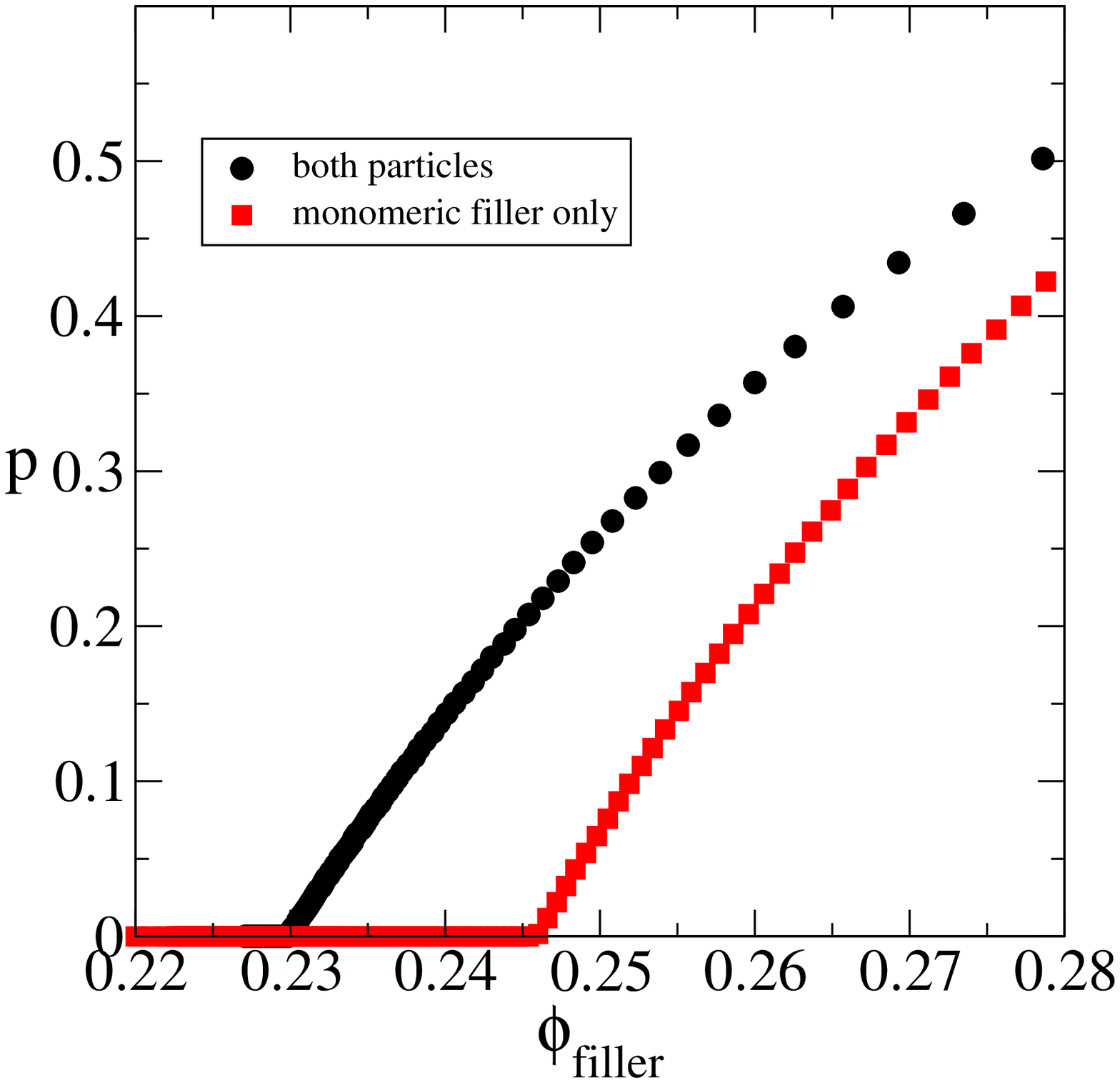}
\end{center}
\par
FIG. 13 \vspace{5cm}
\end{figure}

\newpage

\begin{figure}[tbp]
\begin{center}
\includegraphics[width=8.5cm,angle=270]{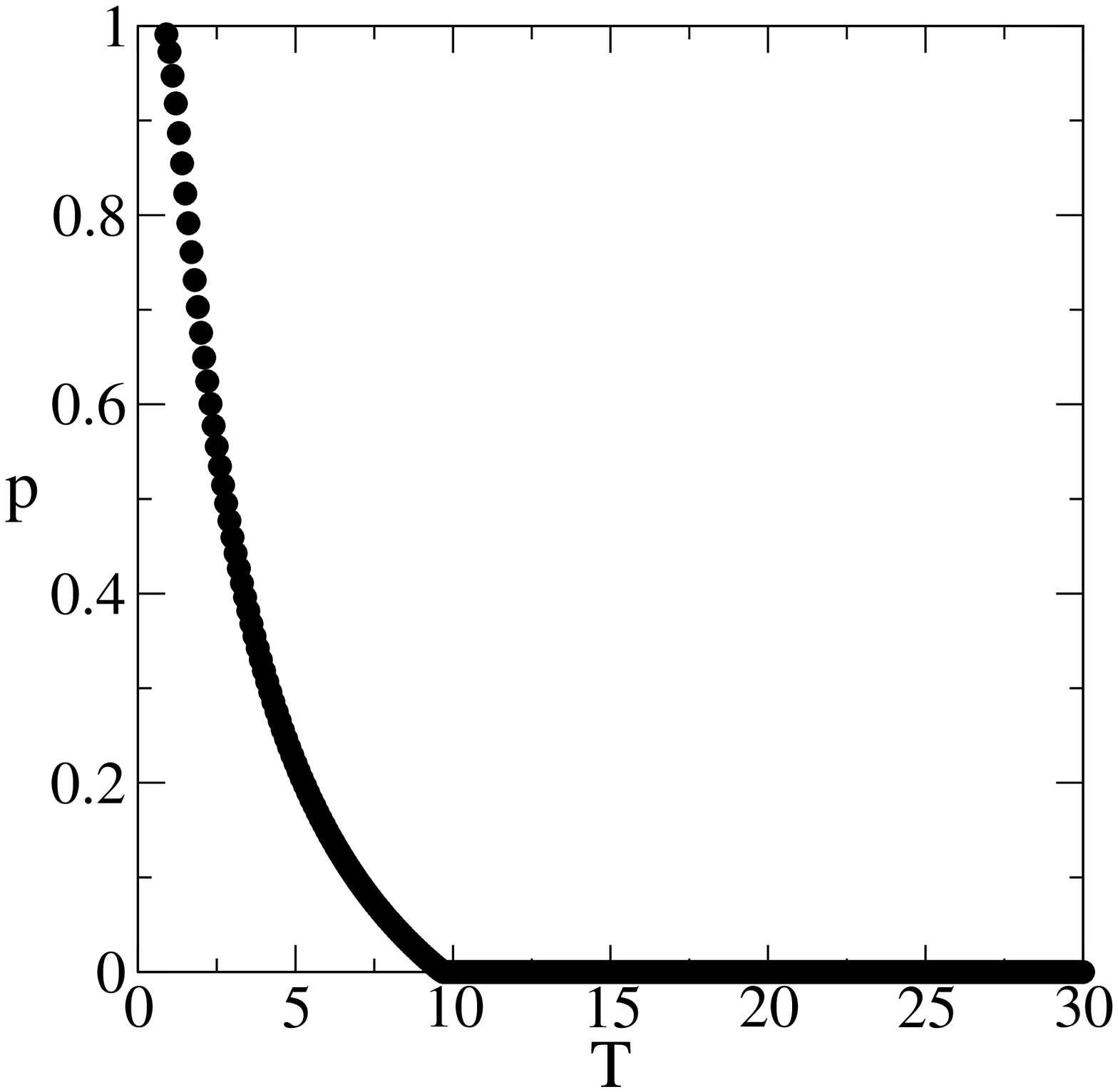}
\end{center}
\par
FIG. 14 \vspace{5cm}
\end{figure}

\newpage
\clearpage

\begin{figure}[tbp]
\begin{center}
\includegraphics[width=8.5cm,angle=270]{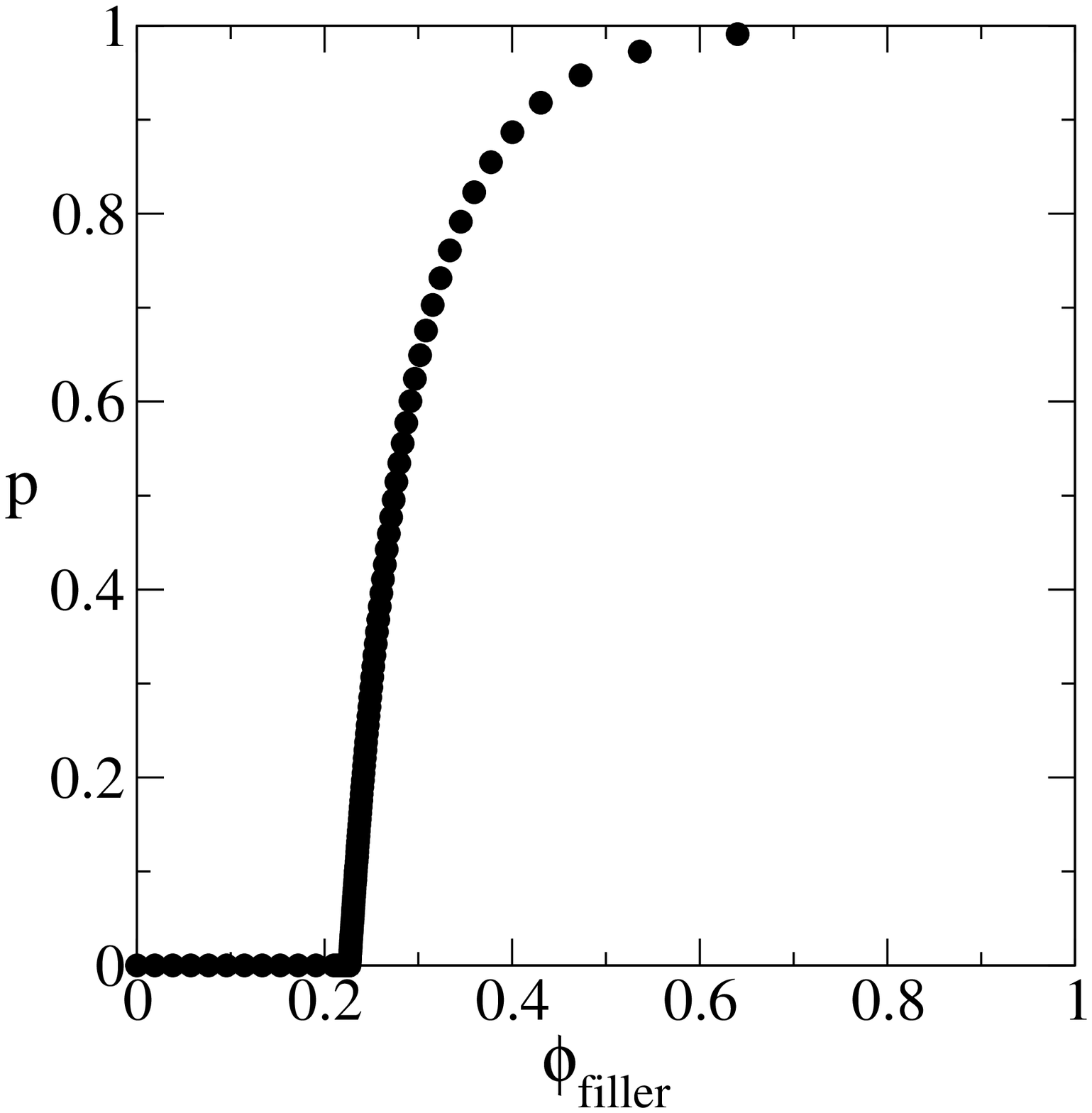}
\end{center}
\par
FIG. 15 \vspace{5cm}
\end{figure}

\newpage

\begin{figure}[tbp]
\begin{center}
\includegraphics[width=8.5cm,angle=270]{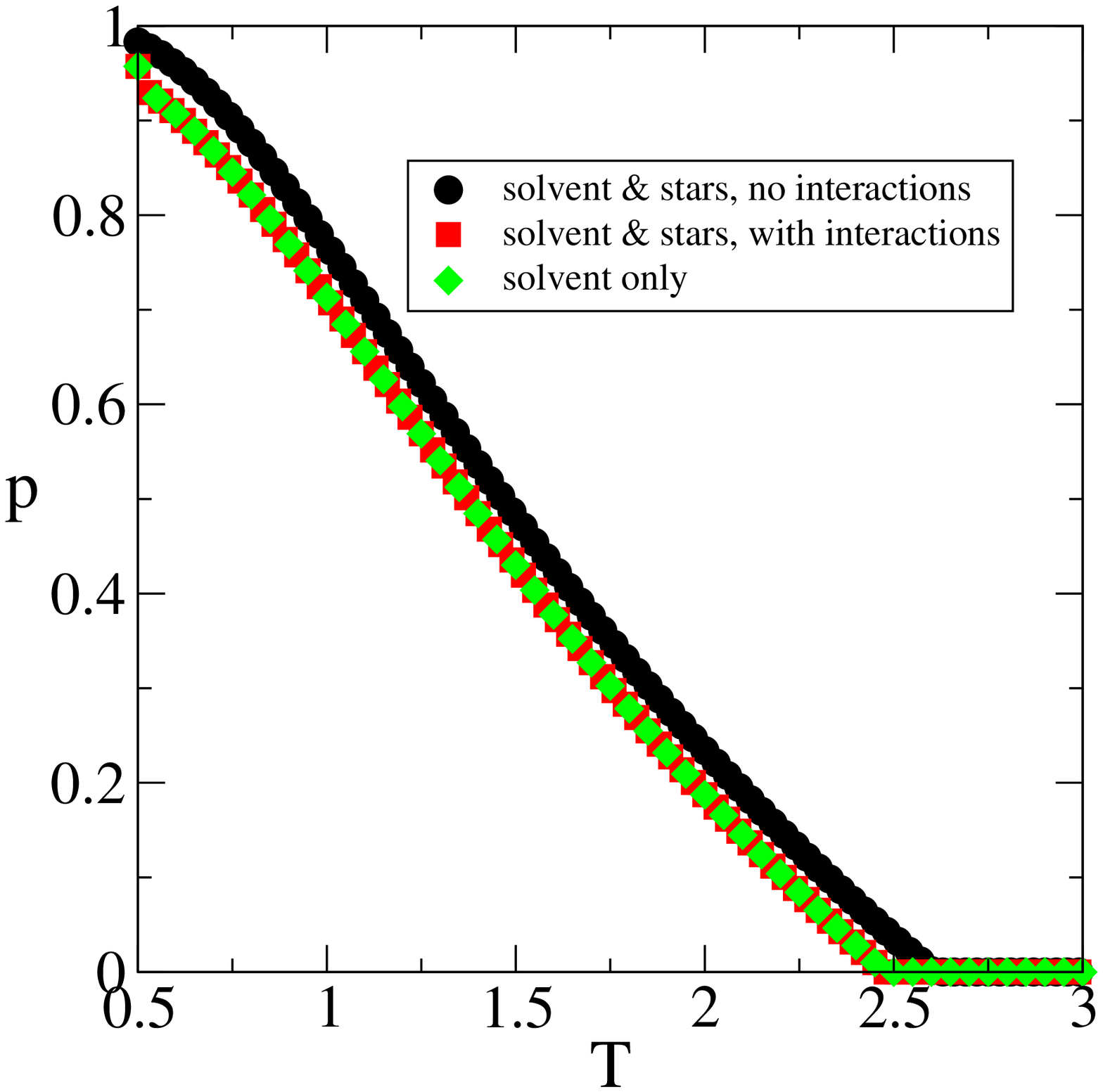}
\end{center}
\par
FIG. 16 \vspace{5cm}
\end{figure}

\newpage

\begin{figure}[tbp]
\begin{center}
\includegraphics[width=8.5cm,angle=270]{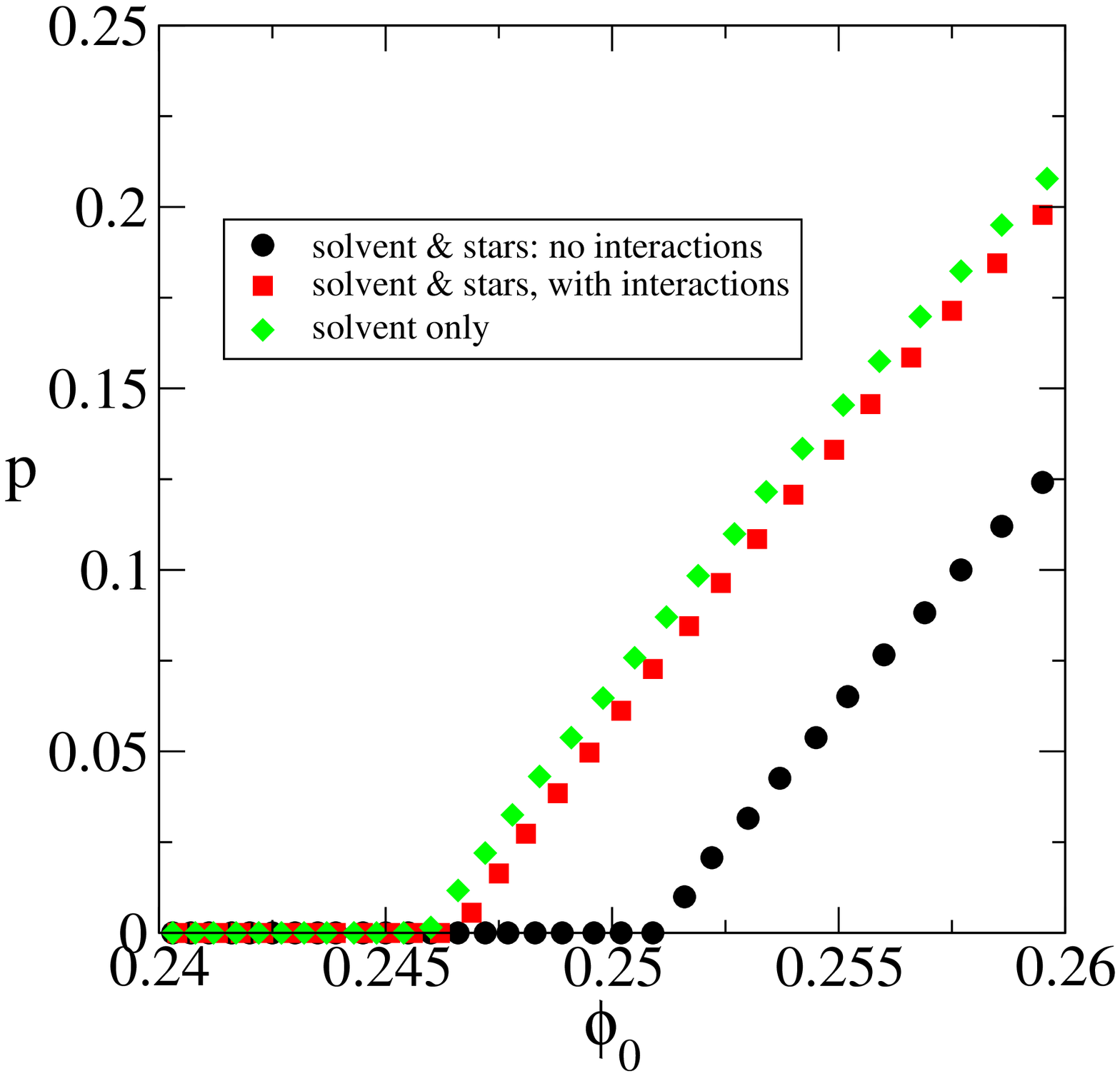}
\end{center}
\par
FIG. 17 \vspace{5cm}
\end{figure}

\newpage

\begin{figure}[tbp]
\begin{center}
\includegraphics[width=8.5cm,angle=270]{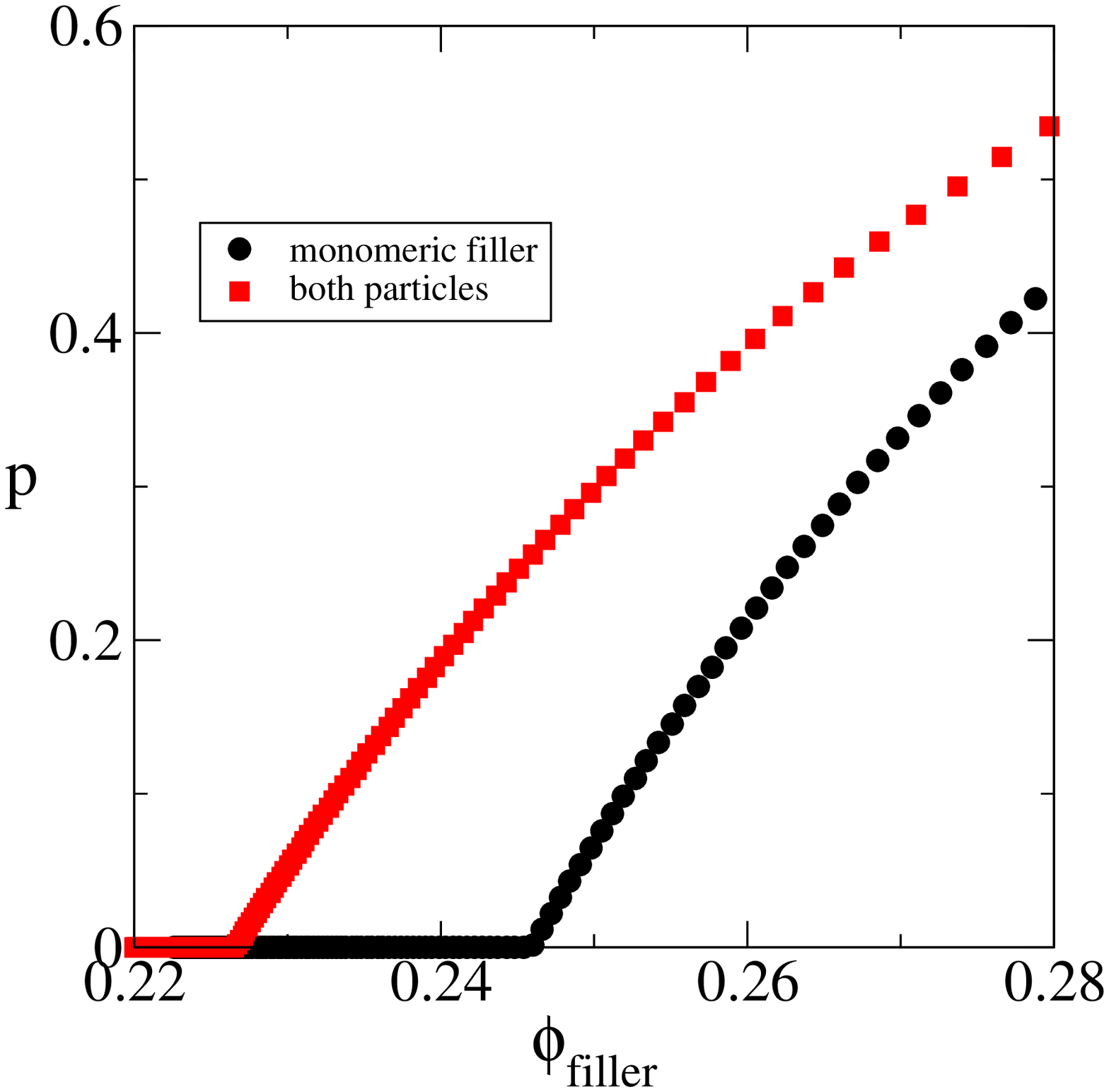}
\end{center}
\par
FIG. 18 \vspace{5cm}
\end{figure}

\newpage

\begin{figure}[tbp]
\begin{center}
\includegraphics[width=8.5cm,angle=270]{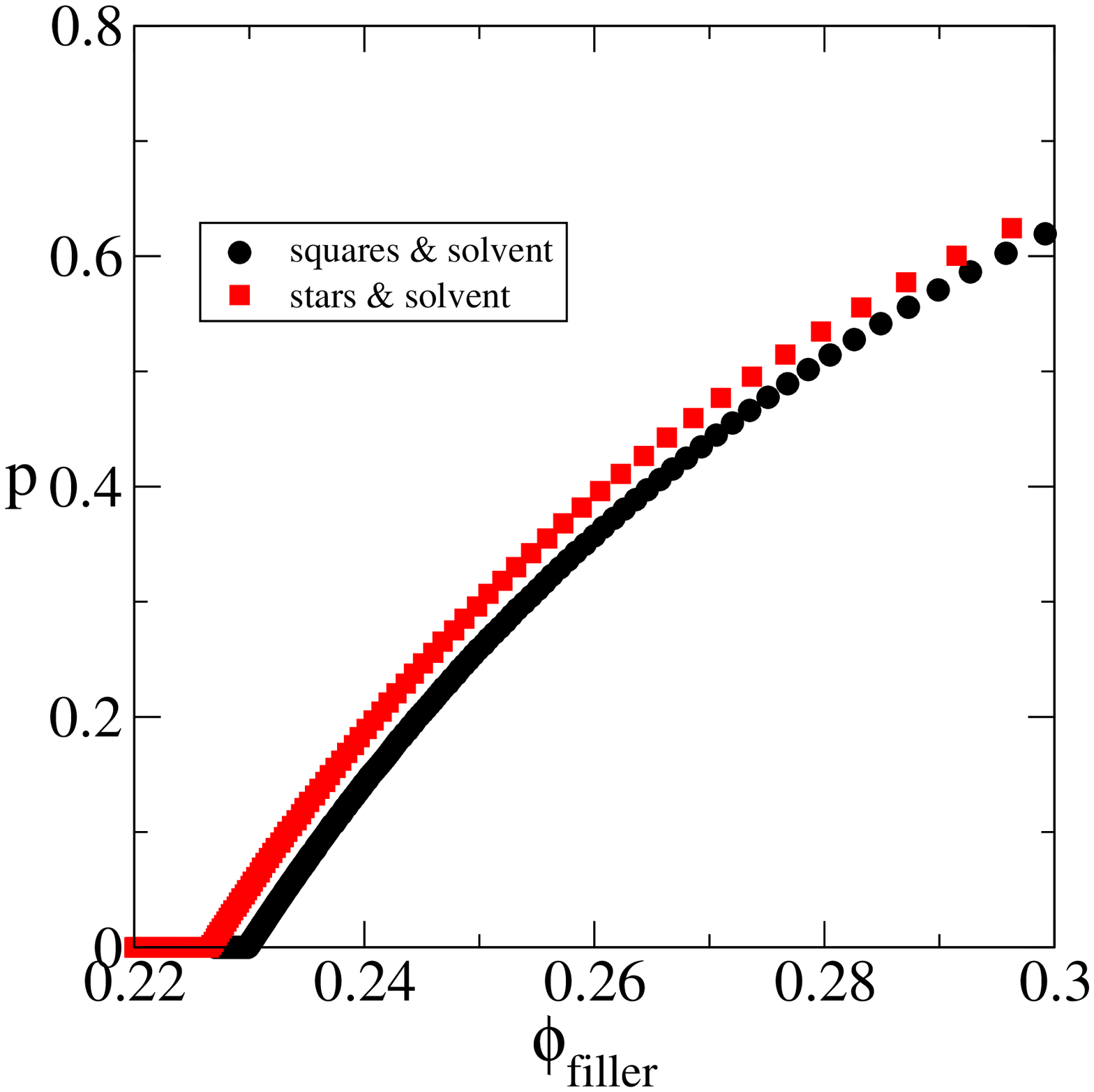}
\end{center}
\par
FIG. 19 \vspace{5cm}
\end{figure}

\end{document}